\documentclass[11pt]{article}
\usepackage{graphicx,tabularx,epsfig}
\usepackage{color}
\usepackage{enumitem}
\usepackage{caption}
\usepackage{subcaption}
\usepackage{amsmath}
\usepackage{amssymb}
\usepackage{amsthm}
\usepackage{physics}
\usepackage{dsfont, mathtools}
\usepackage{psvectorian}
\usepackage{blkarray}
\usepackage{bm}
\usepackage{url}
\usepackage{setspace}

\usepackage{xcolor}
\colorlet{linkequation}{blue}

\usepackage[colorlinks = true,
            linkcolor = blue,
            urlcolor  = blue,
            citecolor = blue,
            anchorcolor = blue]{hyperref}

\usepackage{floatrow}

\newlength{\abstractwidth}
\renewcommand{\thefootnote}{\fnsymbol{footnote}}
\renewcommand{\thanks}[1]{\footnote{#1}} % Use this for footnotes
\newcommand{\starttext}{
\setcounter{footnote}{0}
\renewcommand{\thefootnote}{\arabic{footnote}}}

\makeatletter
\g@addto@macro\normalsize{%
  \setlength\abovedisplayskip{15pt}
  \setlength\belowdisplayskip{15pt}
  \setlength\abovedisplayshortskip{15pt}
  \setlength\belowdisplayshortskip{15pt}
}
\makeatother

\usepackage{color}

\renewcommand{\title}[1]{\vbox{\center\LARGE{#1}}\vspace{5mm}}
\renewcommand{\author}[1]{\vbox{\center#1}\vspace{5mm}}

\begin{document}

\singlespacing

\begin{center}

{\Large \bf {Collisions of localized shocks and quantum circuits}}

\bigskip \noindent
\bigskip
\bigskip

 {\bf Felix M.\ Haehl$^a$ and Ying Zhao$^b$}

\bigskip
\bigskip

    $^a$Institute for Advanced Study, 1 Einstein Dr, Princeton, NJ 08540, USA \vskip1em
    $^b$Kavli Institute for Theoretical Physics, Santa Barbara, CA 93106, USA

\bigskip
\bigskip
    
    {\tt haehl@ias.edu, zhaoying@kitp.ucsb.edu}

\bigskip
\bigskip
\bigskip

\end{center}

\begin{abstract}

We study collisions between localized shockwaves inside a black hole interior. We give a holographic boundary description of this process in terms of the overlap of two growing perturbations in a shared quantum circuit. The perturbations grow both exponentially as well as ballistically. Due to a competition between different physical effects, the circuit analysis shows dependence on the transverse locations and exhibits four regimes of qualitatively different behaviors. On the gravity side we study properties of the post-collision geometry, using exact calculations in simple setups and estimations in more general circumstances. We show that the circuit analysis offers intuitive and surprisingly accurate predictions about gravity computations involving non-linear features of general relativity.

\medskip
\noindent
\end{abstract}

\newpage

\starttext \baselineskip=17.63pt \setcounter{footnote}{0}

{\hypersetup{hidelinks}
\tableofcontents
}

\section{Introduction}

Two uncoupled holographic CFTs can be prepared in an entangled state, which has a dual gravitational description in terms of an Einstein-Rosen (ER) bridge connecting two black hole spacetimes \cite{Maldacena:2001kr,VanRaamsdonk:2010pw,Maldacena:2013xja}. This correspondence is often referred to as ``ER = EPR". In the gravity description, two signals sent from each of the CFTs can meet in the interior once they enter each of the black holes. However, from the boundary point of view, there are no direct interactions coupling the two CFTs, so what boundary feature governs the laws according to which the two signals interact? This puzzling aspect of the correspondence was emphasized in \cite{Marolf:2012xe}.\footnote{ Throughout this paper we assume that the connected two-sided black hole geometry is dual to two disconnected CFTs in an entangled state and that the horizons are smooth as predicted by general relativity. See, however, \cite{Avery:2013bea,Mathur:2014dia} for another perspective.}

 It was pointed out that in the holographic duality, the bulk geometry reflects the minimal tensor network preparing the state \cite{Swingle:2009bg}. In particular, a black hole interior corresponds to a unitary quantum circuit acting on a large number of qubits, c.f., figure \ref{thermofield_double} \cite{Hartman:2013qma,Susskind:2014moa}. Perturbations on the black hole can be represented as perturbations on the circuit evolution, which spread throughout the circuit \cite{Stanford:2014jda,Susskind:2014jwa,Roberts:2014isa}.

In \cite{Zhao:2020gxq}, we gave a quantum circuit interpretation of the `meeting' of two signals sent from different boundaries. We showed that the gravitational interaction between two signals can be encoded in the properties of a quantum circuit with overlapping perturbations. In \cite{Haehl:2021prg,Haehl:2021tft} we proposed boundary quantities that can diagnose the happening as well various properties of the collision process. In particular, we identified the `healthy' (unperturbed) gates as being associated with the post-collision region. These discussions were carried out in the context of spherically symmetric perturbations on AdS-size black holes and only involve the exponential growth of the perturbation (figure \ref{thermofield_double_1}). 

\begin{figure} 
\begin{center}                      
  \includegraphics[width=5in]{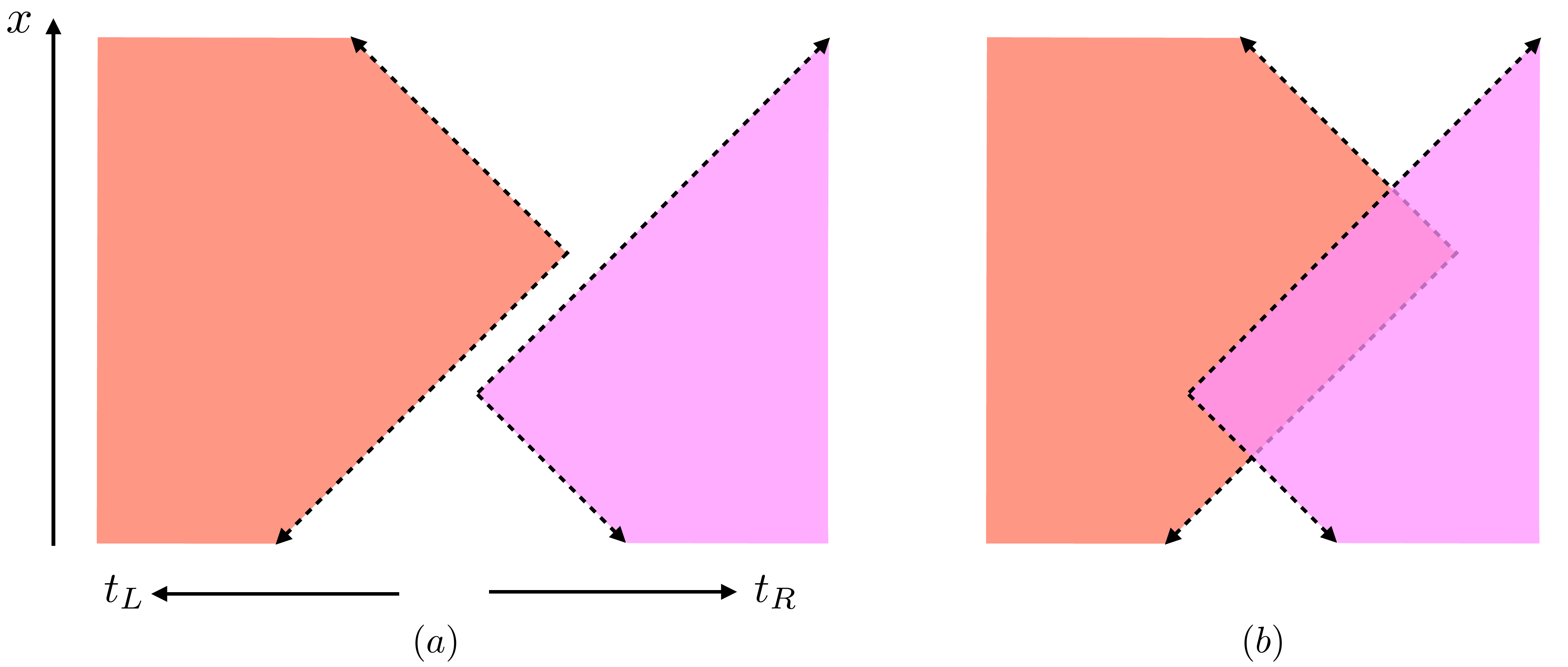}\vspace{-.4cm}
   \caption{The lightcones of two localized perturbations spreading in opposite directions of a quantum circuit may or may not overlap.}
 \label{fig_intro}
  \end{center}
\end{figure}

In this paper, we will consider collisions between {\it localized} shockwaves. We will again be able to interpret their gravitational interaction in terms of the quantum circuit. Because the perturbations are localized, the lattice structure of the quantum circuit plays an important role. Figure \ref{fig_intro} shows a simplified picture of two perturbations ballistically spreading through a quantum circuit with spatial extent. In (a), the two corresponding signals do not feel each other, while in (b), the two signals interact gravitationally. In reality, the perturbations will spread both exponentially as well as ballistically, and the overlapping structure is much more complicated. We will study this in detail and demonstrate a good match between the circuit analysis and gravity computations in general relativity.

Let us briefly mention some other related investigations.
The exponential spreading of perturbations in the quantum circuit is a manifestation of information scrambling, or the quantum butterfly effect \cite{Sekino:2008he,Lashkari:2011yi,Shenker:2013pqa,Maldacena:2015waa}. One way to characterize this phenomenon is in terms of the `growth' of operators \cite{Roberts:2018mnp,Lucas:2018wsc,Qi:2018bje,Magan:2020iac,Jian:2020qpp,Haehl:2021emt,Caputa:2021ori} and their operator complexity \cite{Parker:2018yvk,Magan:2018nmu,Barbon:2019wsy,Rabinovici:2020ryf,Kar:2021nbm,Caputa:2021sib}. Some other related recent studies of the black hole interior using ideas of operator reconstruction
include \cite{Papadodimas:2012aq,Faulkner:2017vdd,Cotler:2017erl,Yoshida:2018ybz,Yoshida:2019kyp,Jafferis:2020ora,Gao:2021tzr}, \cite{Nomura:2018kia,Nomura:2019qps,Nomura:2019dlz,Langhoff:2020jqa,Nomura:2020ska}, and \cite{Leutheusser:2021qhd,Leutheusser:2021frk,Witten:2021unn}.

The paper is organized as follows: in section \ref{sec:sphSymm}, we review and extend previous results on collisions of spherically symmetric shockwaves in AdS-scale black holes. In section \ref{sec:localized} we turn to localized collision and their corresponding quantum circuit description. We show that the behavior can be characterized by four different regimes. We end with a discussion of open problems in section \ref{sec:discussion}. Technical details and some generalizations can be found in appendices.

\subsubsection*{Conventions and terminology}
We list various conventions used in this paper:
\begin{itemize}
     \item All times in are measured in units of thermal time $\frac{\beta}{2\pi}$. We leave this thermal factor implicit so the times are all dimensionless.
    \item $d$ is the spacetime dimension of the boundary theory, so the dual bulk theory has spacetime dimension $d+1$.
    \item $\tilde r_h$ is the horizon radius of the post-collision region. In general, it can have $x$-dependence.
    \item $r_c(x)$ is the radial location of the collision. We call $(u_0(x), v_0(x))$ its coordinate in the unperturbed spacetime, and $(\tilde u_0(x), \tilde v_0(x))$ its coordinates in the post-collision region.
    \item $|x-x_{wL,wR}|$ denotes the proper distances between two points on a one-dimensional line or higher dimensional hyperbolic space $\mathbb{H}_{d-1}$.
\end{itemize}

\section{Collisions of spherically symmetric shockwaves in AdS-size black holes}
\label{sec:sphSymm}

In this section, we review and extend the discussions of collisions between spherically symmetric perturbations in AdS-size black holes and the dual quantum circuit analysis. We mostly follow \cite{Zhao:2020gxq}, but we will introduce some new features and also do calculations in a different way in order to emphasize the essential features that we will use later in the case of localized shocks.

\subsection{Collisions of shockwaves with different scrambling times}
\label{sec:review}

\begin{figure} 
 \begin{center}                      
      \includegraphics[width=.76\textwidth]{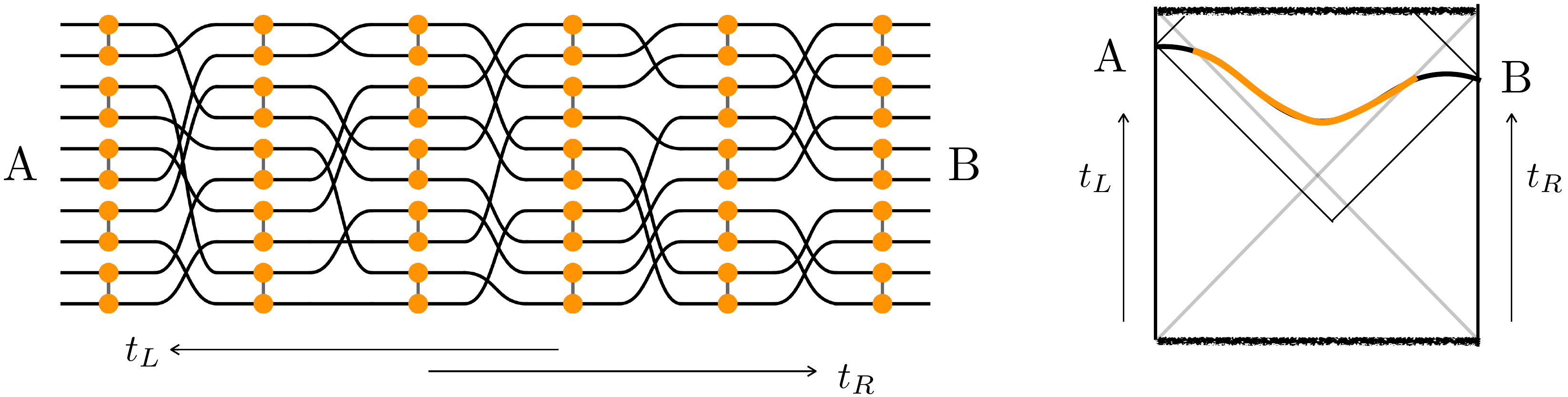}\vspace{-.3cm}
      \caption{{\it Left:} A circuit of quantum gates (e.g., unitary two-qubit operations) describing the evolution of the entangled state between systems A and B. {\it Right:} The AdS wormhole geometry dual to the entangled thermofield double state. The shared quantum circuit represents evolution of the state in the shared interior region.}
  \label{thermofield_double}
  \end{center}
\end{figure}

In earlier work \cite{Zhao:2020gxq,Haehl:2021prg,Haehl:2021tft}, we gave a quantum circuit interpretation of the collision of matter excitations in gravity. It was previously argued that the bulk provides a geometric description of the minimal quantum circuit preparing the boundary state \cite{Swingle:2009bg,Hartman:2013qma,Susskind:2014moa} and bulk regions grow as the number of quantum gates in the circuit grows \cite{Stanford:2014jda,Brown:2015bva,Brown:2015lvg}. When two non-interacting subsystems are entangled with each other, certain quantum gates of the circuit can be `undone' from either side. In this sense these gates are shared between the two subsystems \cite{Zhao:2017isy}. In figure \ref{thermofield_double}, the orange gates can be undone from either subsystem A and subsystem B, and we call them healthy gates. Notice that the growth of the number of healthy gates corresponds to the growth of the interior spacetime region `shared' between A and B, i.e., the region outside the entanglement wedge of either A or B alone but inside the entanglement wedge of A$\,\cup\,$B.

\begin{figure} 
 \begin{center}                      
      \includegraphics[width=.7\textwidth]{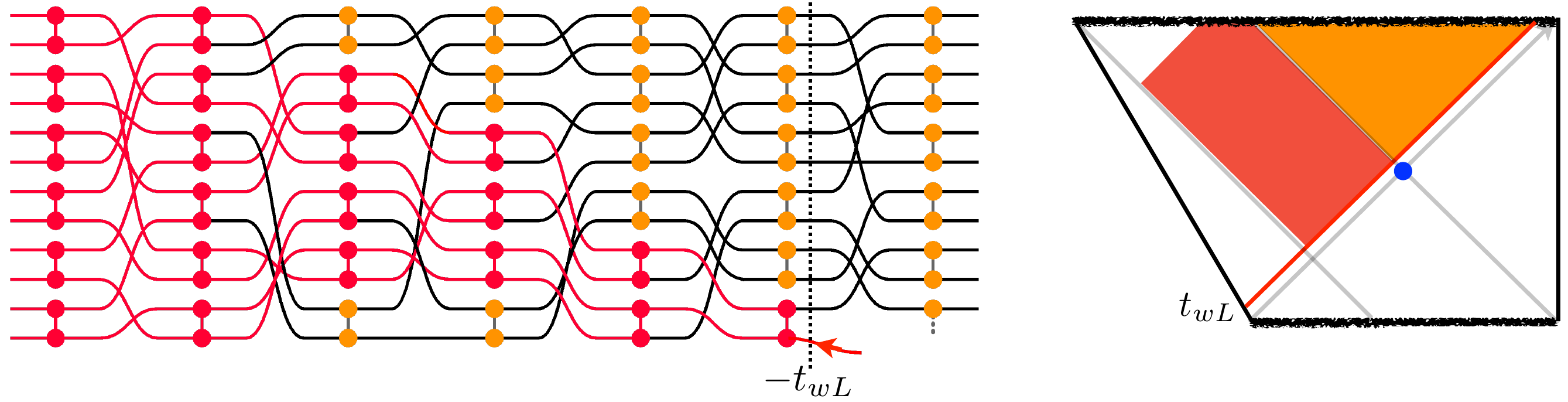}\vspace{-.3cm}
      \caption{{\it Left:} Shared quantum circuit with a perturbation whose effect propagates towards the left, starting at time $-t_{wL}$. {\it Right:} In gravity, the perturbation corresponds to a high energy shockwave, which displaces the left horizon.} 
  \label{thermofield_double_1}
  \end{center}
\end{figure}

When we apply a boundary operator, it creates a perturbation that grows in the quantum circuit. We represent the perturbation applied on the left boundary as the extra red line in figure \ref{thermofield_double_1}. Note that as the perturbation takes time to grow, some healthy gates remain even after the perturbation enters the circuit (orange gates in figure \ref{thermofield_double_1} to the left of $-t_{wL}$). These gates correspond to the shared part of interior region to the future of the shockwave \cite{Zhao:2017isy}.

\begin{figure} 
 \begin{center}                      
      \includegraphics[width=\textwidth]{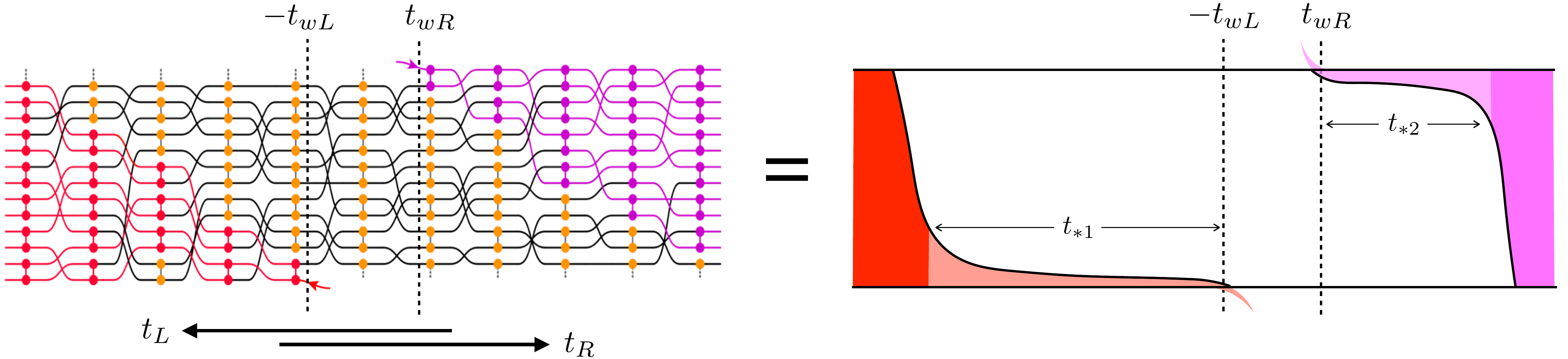}
      \caption{{\it Left:} Quantum circuit with two perturbations (red and pink), which act on the two non-interacting systems (left and right). They affect other qubits through random unitary interaction gates as one proceeds forward in the respective time direction. {\it Right:} schematic picture showing the scrambling times required for each perturbation to grow large as a change in color shade.}
  \label{fig:localizedIntro}
  \end{center}
\end{figure}

 When we apply two operators, one from the left boundary and one from the right boundary, the two operators grow in opposite directions in the quantum circuit. Depending on the relative time the operators were applied, the two perturbations may or may not have overlaps, i.e., there may or may not be an interval of circuit where the two perturbations appear simultaneously. E.g., figure \ref{fig:localizedIntro} shows the case $t_{wR}> -t_{wL}$, where the perturbations have no overlap.
 
In \cite{Zhao:2020gxq} it was argued that when the two growing perturbations overlap in the quantum circuit, the signals collide in the interior. The healthy (orange) gates in the overlap region of the quantum circuit represent the post-collision region in the bulk. In that analysis it was assumed that both perturbations have size of order $1$, i.e., only a small number of qubits are initially affected and the two scrambling times are identical. For the purpose of later applications, we will consider the case where the scrambling times for the two perturbations are different. We achieve this by considering different initial sizes: we denote the number of initially affected qubits as $s_1 = {\cal O}(1)$ while $s_2 = {\cal O}(\sqrt{S})$. With this setup, the scrambling time for the second perturbation will be one half of the scrambling time for first perturbation: $t_{*1} \approx 2 t_{*2}$. 

\begin{figure} 
 \begin{center}                      
      \includegraphics[width=\textwidth]{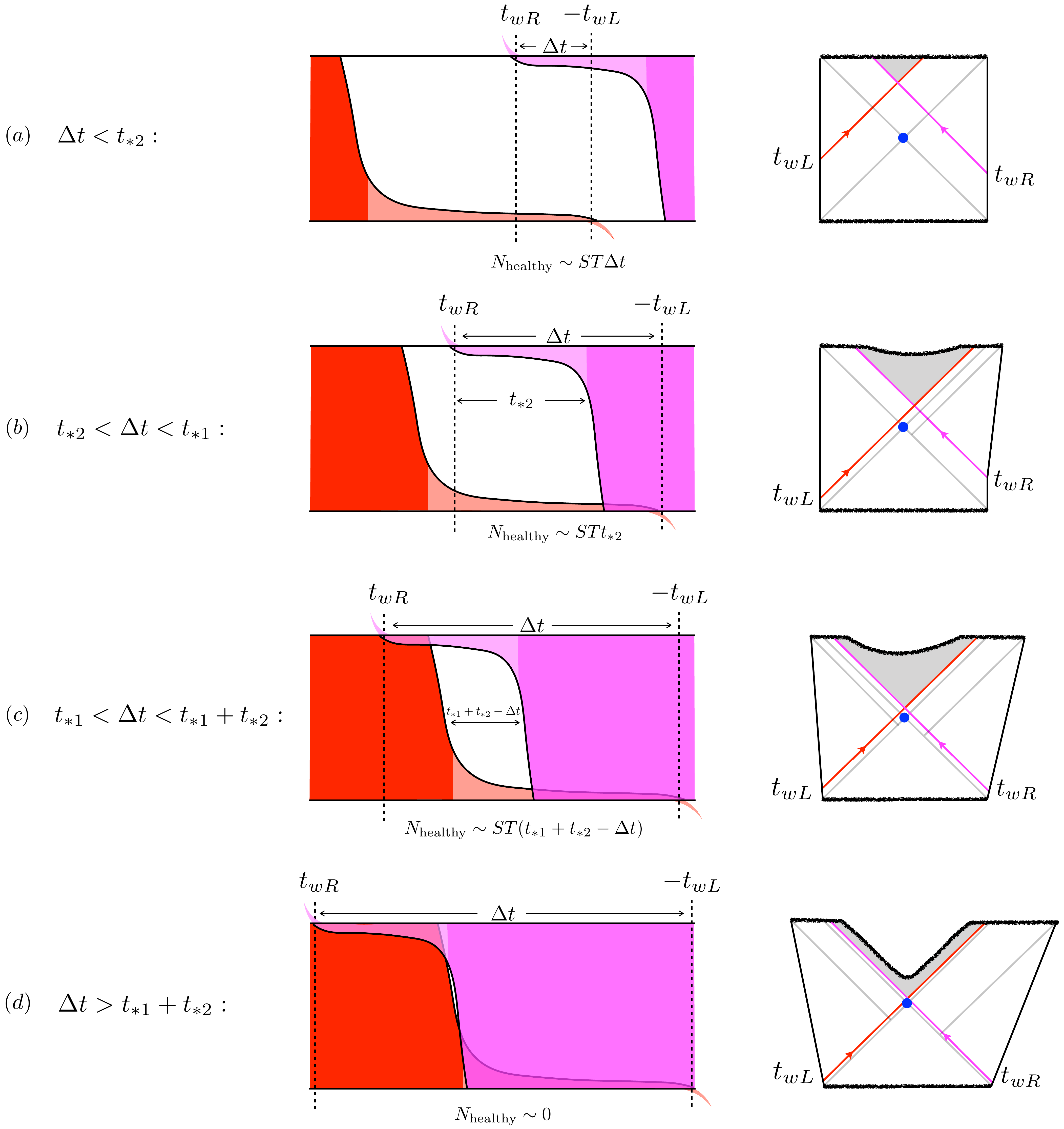}
      \caption{Structure of the quantum circuit and the shockwave geometries as a function of $\Delta t = -t_{wL} - t_{wR}$. We distinguish four qualitatively different regimes, where the dependence of the number of unaffected gates within the shared region of the circuit grows in different ways as a function of $\Delta t$. See main text for details.}
  \label{fig:localizedSlices}
  \end{center}
\end{figure}

In figure \ref{fig:localizedSlices} we illustrate how the perturbations overlap both in the circuit and in the dual eternal black hole geometry, as a function of $\Delta t = - t_{wL} - t_{wR}$. We now describe the four different regimes shown. We are particularly interested in the number of healthy quantum gates (i.e., gates unaffected by both perturbations) in the overlaping part of the circuit and in the volume of the post-collision region:
\begin{enumerate}
\item In figure \ref{fig:localizedSlices}(a), we assume $\Delta t<t_{*2}$. In this case, neither of the perturbations has grown large in the overlap region and the number of healthy gates in the overlap region roughly grows linearly in $\Delta t$. In the dual AdS black hole picture, the collision point moves closer to the bifurcating surface and the post-collision region (grey in the figure) grows as $\Delta t$ increases. 

\item In figure \ref{fig:localizedSlices}(b), we show the case when $t_{2*}<\Delta t<t_{*1}$. In this case, the number of healthy gates in the overlap stays approximately constant as it cannot be larger than $ST t_{*2}$. In the gravity picture, as the collision energy increases, there is significant backreaction, which causes the singularity to bend down. This reduction of post-collision spacetime volume compensates for its increase due to the fact that the collision occurs earlier. 

\item In figure \ref{fig:localizedSlices}(c), we consider $t_{1*}<\Delta t<t_{*1}+t_{*2}$. In this case, the number of healthy gates in the overlap region starts to decrease linearly in $\Delta t$. In the bulk, the collision point is very close to the horizon. Backreaction causes the interior to shrink.

\item In figure \ref{fig:localizedSlices}(d), we illustrate the case when $\Delta t>t_{*1}+t_{*2}$. At this point, there are almost no healthy gates in the circuit as the entire overlap region is covered by at least one perturbation that has grown to its maximum size. The singularity has now bent down significantly such that the volume of the post-collision region has shrank to an exponentially small size.
\end{enumerate}

In gravity, we thus see that a competition of two effects gives rise to four regimes.

\paragraph{Detailed comparison in perturbed BTZ geometry:} 
We can make our observations quantitatively more precise through explicit computations in the BTZ black hole with spherically symmetric perturbations. In this case the geometry of the post-collision region is exactly known \cite{Dray:1985yt,Shenker:2013yza} and we can compute the time dependence of the spacetime volume. We match this gravitational time dependence to the epidemic model of operator growth \cite{Susskind:2014jwa}.

\begin{figure}
 \begin{center}                      
      \includegraphics[width=.49\textwidth]{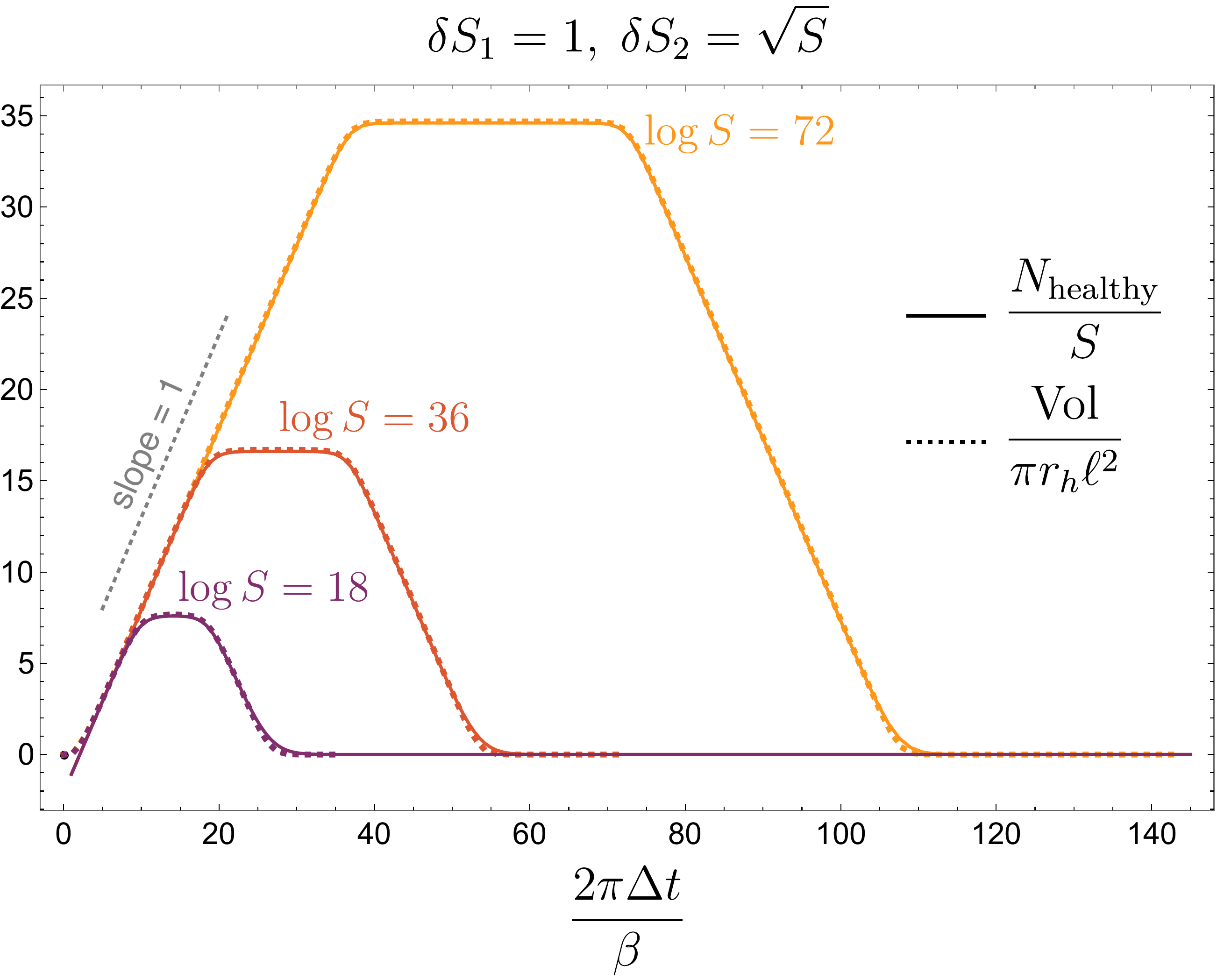}
      \includegraphics[width=.49\textwidth]{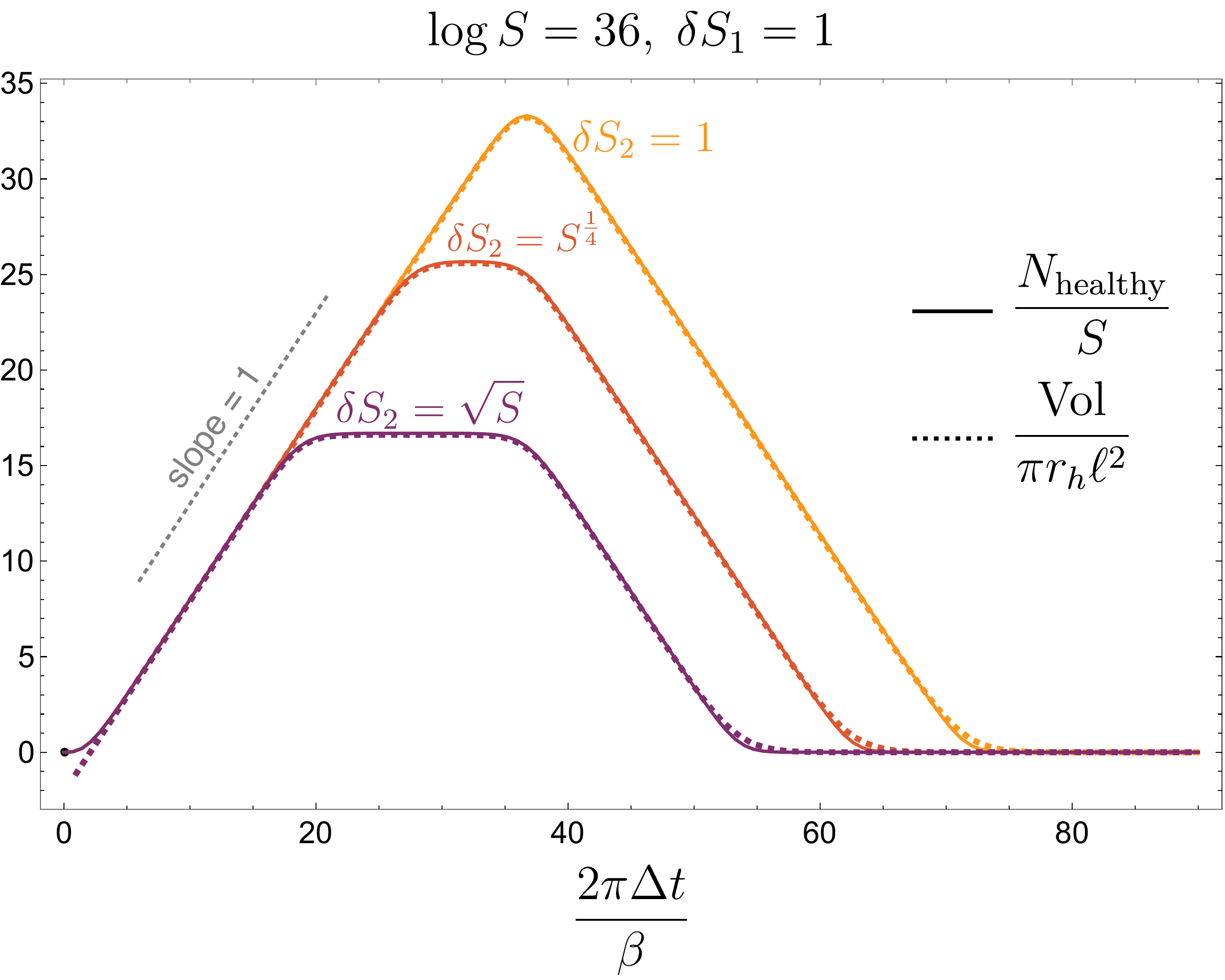}
      \vspace{-.4cm}
      \caption{{\it Left:} Comparison of post-collision spacetime volume and number of `healthy' gates in the circuit with two perturbations with initial `size' $\delta S_1=1$ and $\delta S_2 = \sqrt{S}$.\protect\footnotemark $\,$ We can clearly see the four regimes discussed in the text. They get more pronounced for larger values of $\log S$. {\it Right:} The same quantities for fixed $\log S = 36$ and different values of $\delta S_2$. The plateau only exists for sufficiently different scrambling times.}
  \label{plot_1}
  \end{center}
\end{figure}

In figure \ref{plot_1}, we see the four different regimes described above for both the number of healthy gates in the overlap region and the volume of the post-collision region: linear growth, constant `balance', linear decrease, and exponentially small late time behavior. \footnotetext{The epidemic model for the quantum circuit with two perturbations uses as input parameters the initial sizes $s_i$ of the perturbations $i=1,2$. These are model parameters, which we have empirically fitted to give the best possible match with the more stringent gravity result. We find that the ideal value for the initial size of the circuit perturbations is four times the change of entropy in gravity: $s_i^{\text{(circuit)}} = 4 \,\delta S_i^\text{(gravity)}$. The factor of 4 appears to be robust and independent of any other parameters. We leave a detailed understanding of this observation to the future.} Except for some transient effects (very early time and near $t_{*1}+t_{*2}$), the match between the circuit estimates and the gravity calculation in the BTZ geometry is suprisingly good for all regimes considered.

\subsection{Estimates using the Raychaudhuri equation}
\label{sec:estimate_Raychaudhuri_1}

In the previous subsection we presented detailed results for the BTZ geometry with spherically symmetric perturbations. In more general cases the backreacted metric with two shockwaves is unknown and we need to resort to estimates. We will now present an estimation of the corresponding volume using the Raychaudhuri equation. Specifically, we focus on the third and fourth regimes described above, in order to illuminate the gravitational mechanism leading to a shrinking post-collision spacetime at late times.

\subsubsection{Bound on affine distances}

\begin{figure} 
 \begin{center}                      
      \includegraphics[width=.7\textwidth]{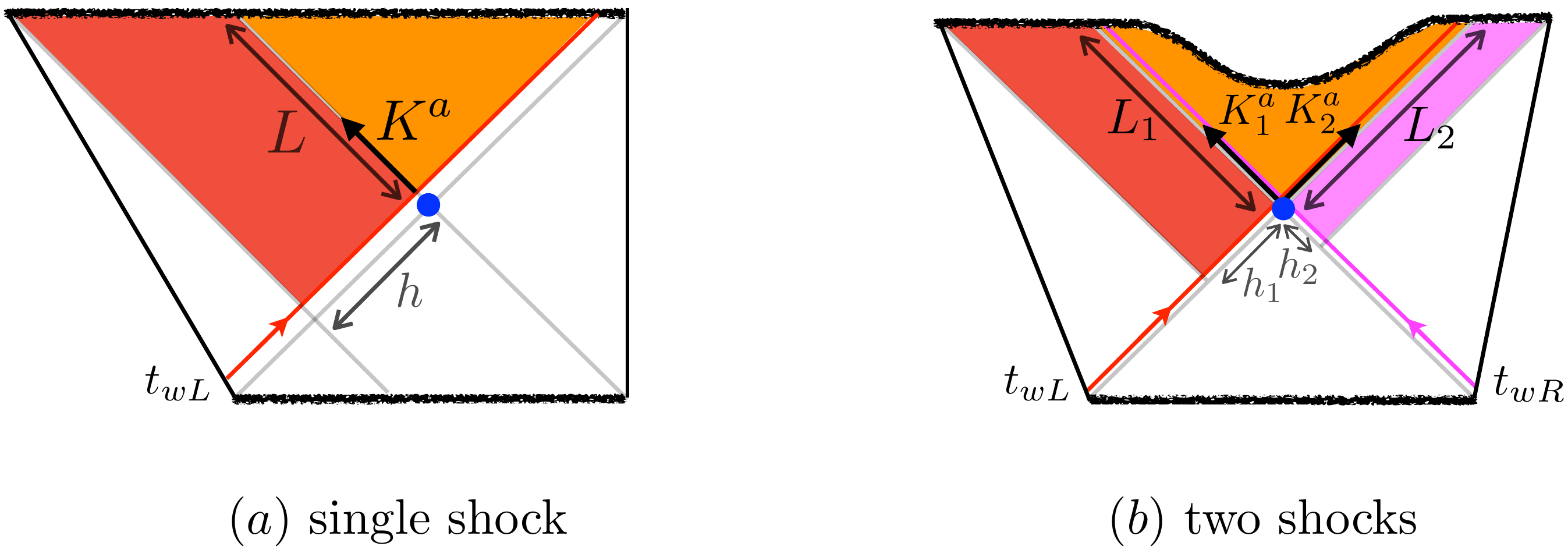}
      \caption{{\it Left:} A single shockwave induces a discontinuity proportional to $h$ in the expansion of a family of radial null geodesics with tangent vector $K$. {\it Right:} With two colliding shocks we consider two null vectors. Raychaudhuri's equation implies an upper bound on the product of affine distances $L_1 L_2$, and hence on the post-collision spacetime volume.}
  \label{Raychaudhuri_1}
  \end{center}
\end{figure}

Consider radial null lines with tangent vector $K$ passing the bifurcating surface in the presence of a single shockwave (left panel of figure \ref{Raychaudhuri_1}). The Raychaudhuri equation characterizes the evolution of the expansion $\theta$ of a congruence of curves \cite{Hawking:1973uf,Landau:1975pou}. In the case of null lines in $d+1$ spacetime dimensions, we have
\begin{align}
\label{Raychaudhuri}
	\dot\theta = -\frac{\theta^2}{d-1}-\sigma^2+\omega^2-R_{ab}K^aK^b+\dot K^a_{;a}
\end{align}
where $\omega^2 = \omega_{ab}\omega^{ab}\geq 0$ and $\sigma^2 = \sigma_{ab}\sigma^{ab}\geq 0$ are the vorticity and shear which cause the congruence of curves to expand and contract, respectively. 

The expansion of radial null geodesics is zero at the bifurcating surface. Assuming the shockwave stays close to the horizon, we can consider the expansion right after the geodesics cross the shockwave. The stress-energy tensor of the shockwave contains a delta function at the horizon (c.f., \cite{Shenker:2013pqa}):
\begin{align}
	R_{ab}K^aK^b = 8\pi G_N T_{ab}K^aK^b = 2h\delta(u)K^uK^u \,,
\end{align}
where we used $g_{ab}K^aK^b = 0$. We infer that across the shockwave, the expansion jumps from $\theta =0$ to a negative number $\theta_1 = -2h K^u(u=0) \equiv -2hC_1$, where $C_1$ is a normalization constant.

Intuitively, once the expansion becomes negative, gravitational contraction will only make it more negative. To see this, we again consider the Raychaudhuri equation, which now implies the following inequalities:\footnote{ Note that we ignore the vorticity term as it vanishes in our case. The last term in \eqref{Raychaudhuri} vanishes for tangent vectors of geodesics. The curvature term is non-negative due to the null energy condition.}
\begin{equation}
\label{bound_0}
	\dot\theta\leq-\frac{\theta^2}{d-1} \qquad \Rightarrow \qquad 
	L_1\leq-\frac{d-1}{\theta_1} \equiv \frac{1}{2h C_1} \,,
\end{equation}
where $L_1$ is the affine time required by the null geodesic to reach the singularity after passing the shockwave (figure \ref{Raychaudhuri_1}). In general, the divergence of expansion does not necessarily imply a singularity. However, in our case we do know from spherical symmetry that the singularity exists. As null affine time is defined up to a constant, this bound is not physically meaningful; it depends on the normalization of the vector $K$ through the constant $C_1$.

Let us now consider the case with two shockwaves (right panel of figure \ref{Raychaudhuri_1}). In this case, a similar argument as above leads to a physically meaningful bound. The details of the corresponding calculation are given in appendix \ref{app:GeoSphSymm}, but we will now provide a summary. We take two null vectors $K_1$ and $K_2$ and normalize them such that $K_1\cdot K_2 = -2\ell^2$, which implies $C_1C_2 = 1$. With this condition we see that the product of the affine distances along the left and right null directions in the right panel of figure \ref{Raychaudhuri_1} satisfies
\begin{align}
\label{bound_1}
	L_1L_2\leq \frac{1}{4h_1h_2C_1C_2} = \frac{1}{4h_1h_2} \,.
\end{align}

\subsubsection{Post-collision geometry}
\label{subsec:post_collision}

The inequality \eqref{bound_1} further provides an estimated bound on the volume of the post-collision region based on some assumptions. In fact, if the geometry in the post-collision region is completely arbitrary, we cannot derive any bound from \eqref{bound_1}. But we do know that the geometry needs to satisfy Einstein's equation in the vacuum with a negative cosmological constant. In $2+1$ bulk dimensions, we also know that it is locally $AdS_3$. We can thus consider a post-collision metric of the following form:
\begin{align}
	ds^2 = -\frac{4\ell^2 du dv}{(1+uv)^2}+\tilde r_h^2\qty(\frac{1-uv}{1+uv})^2 d\phi^2 \,,
\end{align}
where $\tilde{r}_h$ is the horizon radius after the collision.
We take the collision point to be at $(\tilde u_0, \tilde v_0)$ in these coordinates. With $K_1\cdot K_2 = -2\ell^2$, we have 
\begin{align}
    &L_1L_2 = \frac{(1-\tilde u_0\tilde v_0)^2}{4\tilde u_0\tilde v_0} = \frac{1}{\frac{\tilde r_h^2}{r_c^2}-1} \qquad \Rightarrow \qquad \frac{\tilde{r}_h}{r_c} = \sqrt{1+\frac{1}{L_1L_2}}
    \,,
\end{align}
where $r_c$ is the radial location of the collision point.
Using this relation, as well as the relation between collision point and pre-collision horizon radius, $\frac{r_c}{r_h}= \tanh(\frac{\Delta t}{2})$, we find for the spacetime volume of the post-collision region:
\begin{align}
\frac{V}{\pi r_h\ell^2} =  \ &\frac{\tilde r_h}{r_h}\qty[\log(\frac{\frac{\tilde r_h}{r_c}+1}{\frac{\tilde r_h}{r_c}-1})-2\frac{r_c}{\tilde r_h}]\nonumber\\
 =\ & \tanh(\frac{\Delta t}{2})\sqrt{1+\frac{1}{L_1L_2}}\,\log(\frac{\sqrt{1+\frac{1}{L_1L_2}}+1}{\sqrt{1+\frac{1}{L_1L_2}}-1})-2\tanh(\frac{\Delta t}{2})
\label{bound_volume_0}\\
    \leq \ &\tanh(\frac{\Delta t}{2})\sqrt{1+4h_1h_2}\,\log(\frac{\sqrt{1+4h_1h_2}+1}{\sqrt{1+4h_1h_2}-1})-2\tanh(\frac{\Delta t}{2})\nonumber\\
    \approx \ &\begin{cases}
 	t_{*1}+t_{*2}-\Delta t &\quad  \Delta t<t_{*1}+t_{*2}\\
 	\frac{2}{3}e^{-(\Delta t- t_{*1}-t_{*2})} &\quad \Delta t> t_{*1}+t_{*2}
 \end{cases}\label{bound_volume_1}
\end{align}
where we identified the null shift due to the shocks as $2h_1 = \frac{\delta S_1}{S} \, e^{-t_{wL}}$ and similarly for $h_2$ (c.f.\ \cite{Shenker:2013pqa}). We also used $\Delta t\gg 1$.

Note that the approximate upper bound \eqref{bound_volume_1} reproduces the circuit behavior in the late time regimes 3 and 4, where the number of healthy gates decreases linearly in $\Delta t$ and stays exponentially small, respectively. This retrospectively justifies the assumptions made in the above derivation. 

Also note that in equations \eqref{bound_0} and \eqref{bound_1}, we already assumed the shockwave to be close to the horizon $u = 0$ and therefore we only obtained the late time behavior in regimes 3 and 4. As we will show in later sections, one can incorporate the early time behavior by improving the inequalities in \eqref{bound_0} and \eqref{bound_1}.

\section{Collisions of localized shockwaves}
\label{sec:localized}

In this section, we generalize the earlier discussion to the case concerning the collision of two localized shockwaves. We first provide an appropriate circuit picture, and then discuss the four regimes of the growing and shrinking post-collision region in gravity.

\subsection{Circuit analysis}

Consider one localized shock due to the insertion of a local operator. The quantum circuit picture with one localized shock was discussed in \cite{Roberts:2014isa}. In particular, figure 1 of \cite{Roberts:2014isa} inspires our discussion: two `cones' can be associated with the spread of the localized perturbation. A priori, the operator spreads inside the causal light cone, but its effect stays small until it reaches the butterfly cone. The butterfly cone is generally bounded by time-like rays, whose slope is given in terms of the butterfly velocity $v_B$.

\begin{figure} 
 \begin{center}                      
      \includegraphics[width=4.2in]{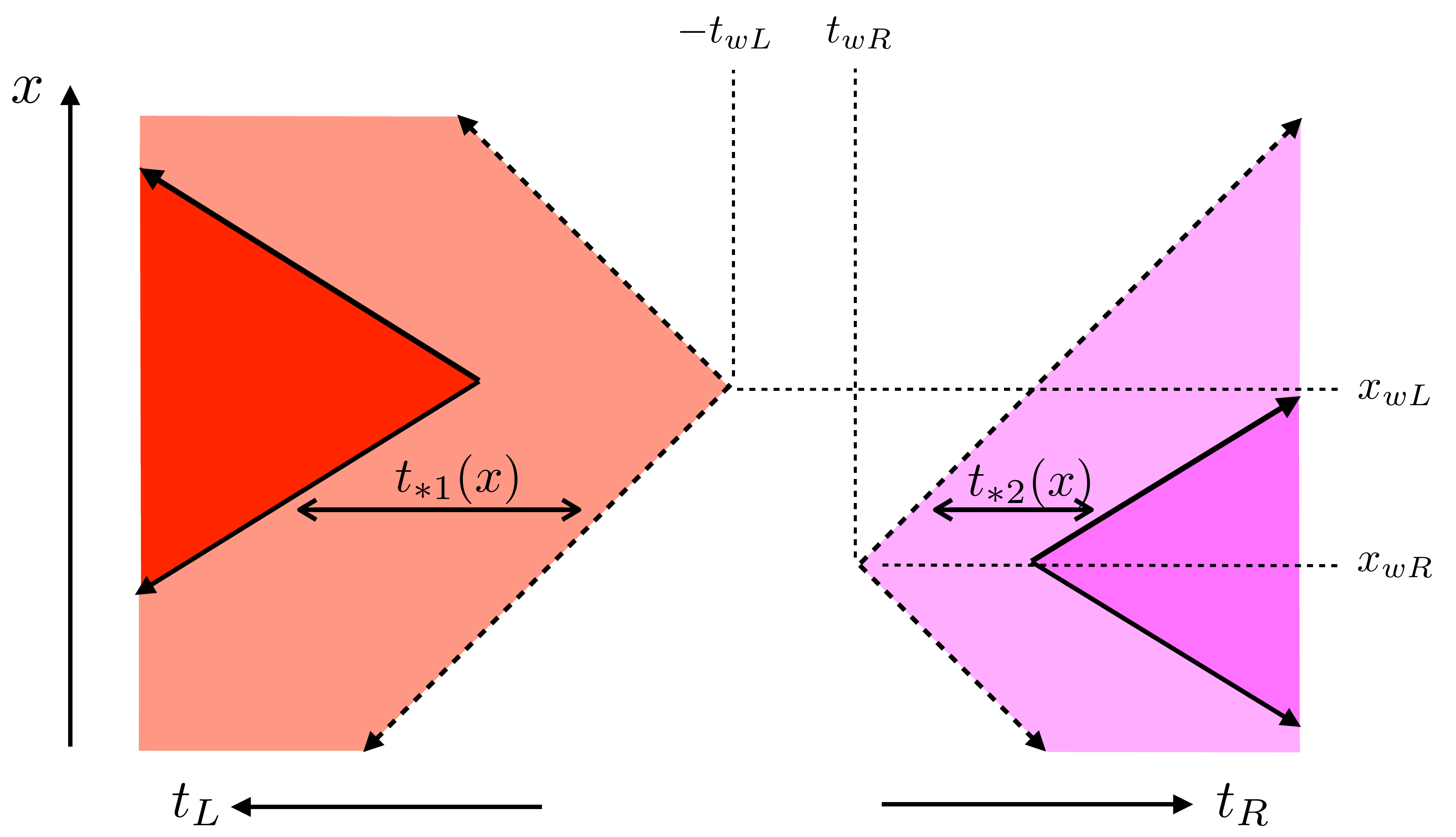}
      \caption{Two localized shocks propagating in the shared quantum circuit. Light colors correspond to the causal light cones, where only few gates are affected by the perturbation. Dark colors correspond to the butterfly cones, where the perturbation has grown significantly. The two types of cones are separated by an $x$-dependent scrambling time.}
  \label{fig:localizedIntro2}
  \end{center}
\end{figure}

For our purpose, we consider two sets of such cones, which describe the growth of two localized perturbations in opposite directions, see figure \ref{fig:localizedIntro2}. As before, the red perturbation propagates toward the left and the pink perturbation propagates toward the right, i.e., they correspond to perturbations of the left and right systems, respectively. The light red (pink) region represents the causal light cone of the left (right) perturbation, while the dark (red) pink region represents the corresponding butterfly cone. As we will argue below, constant-$x$ slices of this picture correspond to quantum circuits such as those illustrated in figure \ref{fig:localizedIntro}.
  
We focus on some particular fixed value of $x$.\footnote{In this paper, by constant $x$ slices we always have in mind slices with a width of roughly $\ell_{AdS}$.} In slight abuse of language, we call the distance between the causal light cone and the butterfly cone as $t_*$, as this is the time it takes for the perturbation to significantly affect operators at location $x$ after they first come in causal contact. Note that this definition of $t_*$ will depend on $x$ when the butterfly velocity is less than $1$. Without loss of generality, we again assume $t_{*1}(x)>t_{*2}(x)$ (figure \ref{fig:localizedIntro2}). We have
 \begin{equation}
     \begin{split}
 	&t_{*1}(x) = \frac{\beta}{2\pi}\log\frac{c}{\delta S_1}+\qty(\frac{1}{v_B}-1)|x-x_{wL}|\,,\\
 	&t_{*2}(x) = \frac{\beta}{2\pi}\log\frac{c}{\delta S_2}+\qty(\frac{1}{v_B}-1)|x-x_{wR}|\,,
 	\end{split}
 	\label{eq:tstarDef}
 \end{equation}
 where $c$ is the central charge of the system.

\begin{figure}
 \begin{center}                      
      \includegraphics[width=.8\textwidth]{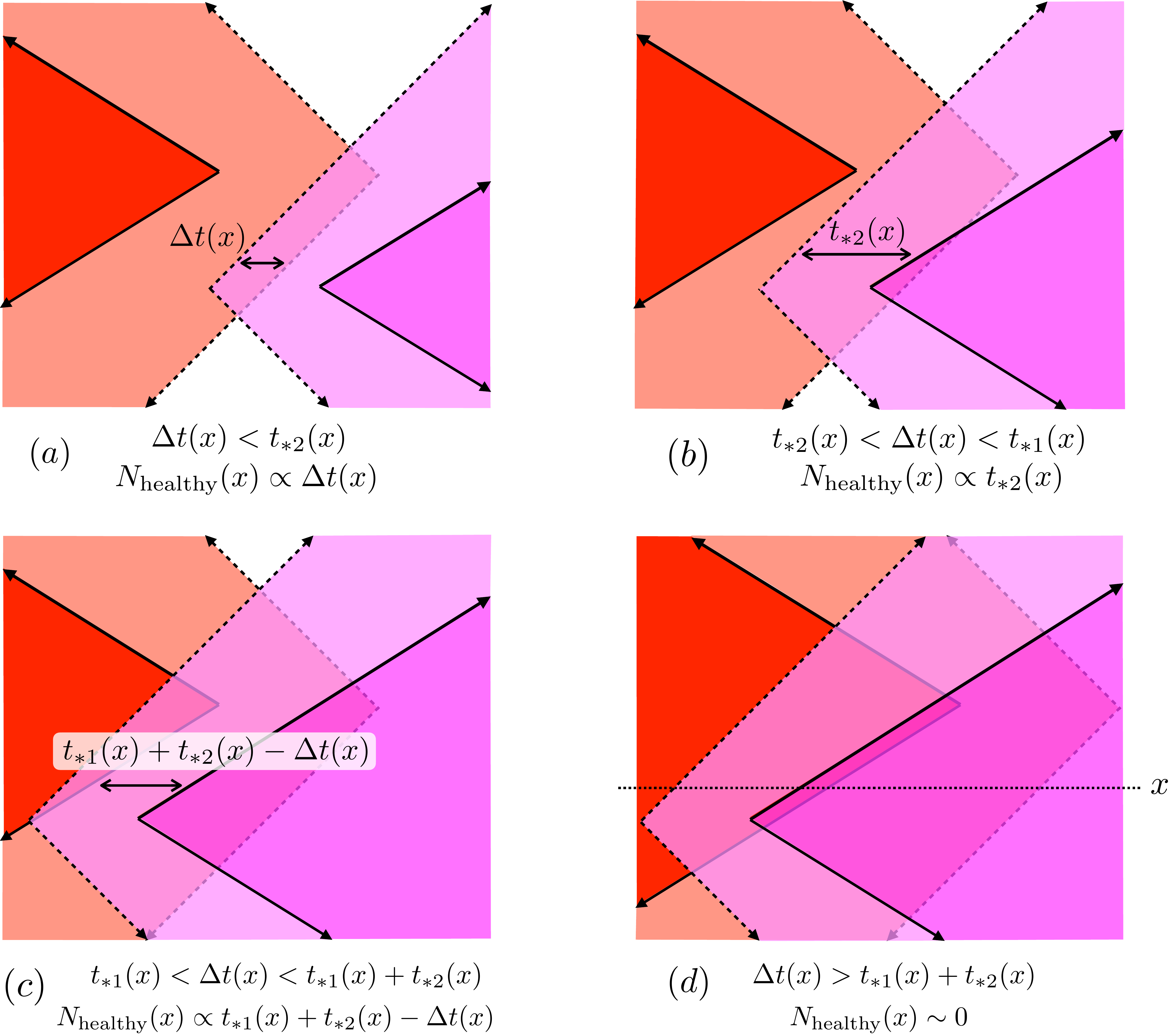}\vspace{-.3cm}
      \caption{We show four dynamical regimes for the number of healthy gates in the overlap region of two localized perturbations. The regimes are in correspondence with those shown in figure \ref{fig:localizedSlices}.}
  \label{fig:localizedSummary}
  \end{center}
\end{figure}

We now turn to a discussion of the same four regimes that also governed the setup described in section \ref{sec:review}. In figure \ref{fig:localizedSummary} we illustrate these four regimes for the case of localized shocks.
Let 
\begin{equation}
\label{eq:DeltatDef}
\Delta t(x) \equiv -t_{wL}-|x-x_{wL}|-t_{wR}-|x-x_{wR}|    
\end{equation}
be the amount of overlap of the two perturbations at transverse location $x$. 
In figure \ref{fig:localizedSummary}(a), we show the case when $\Delta t(x)>0$, i.e., the two perturbations begin to have overlap at transverse location $x$. We call $N_{\text{healthy}}(x)$ the number of healthy gates in the overlap region at transverse location $x$. As long as $\Delta t(x)<t_{*2}(x)$, $N_{\text{healthy}}(x)$ again grows linearly in $\Delta t(x)$. Next, we fix the impact parameter and gradually increase $\Delta t$. When $t_{*2}(x)<\Delta t(x)<t_{*1}(x)$, the number of healthy gates in the overlap region at transverse location $x$ stays constant, see figure \ref{fig:localizedSummary}(b). In figure \ref{fig:localizedSummary}(c), we let $t_{*1}(x)<\Delta t(x)<t_{*1}(x)+t_{*2}(x)$. In this regime, the number of healthy gates in the overlap region at transverse location $x$ is proportional to $t_{*1}(x)+t_{*2}(x)-\Delta t(x)$ and decreases linearly with $\Delta t(x)$. In the late time regime when $\Delta t(x)> t_{*1}(x)+t_{*2}(x)$, we no longer expect there to be healthy gates at transverse location $x$, c.f., figure \ref{fig:localizedSummary}(d).

Note that these four regimes (at constant $x$) are analogous to the four regimes we found in the case of AdS-size black holes with parametrically different scrambling times (e.g., figure \ref{fig:localizedSlices}). We can thus think of each constant $x$ slice in figures \ref{fig:localizedSummary} as being described by the AdS-size black hole analysis. In the following we will make this more precise.

\subsection{Gravity analysis}

On the bulk side, the post-collision geometry is in general unknown. The goal of this subsection is to describe certain general features qualitatively and compare them with the circuit analysis. From the latter we argued for four different regimes. We will now identify different gravitational mechanisms that are responsible for these different behaviors in the case of colliding localized shocks. 

\subsubsection{Growth of the post-collision region}
\label{sec:growth}

At early times, when $\Delta t(x)<t_{*2}(x)$, non-linear effects can be ignored and we may assume that the two shockwaves propagate freely in the original black hole geometry. To quantify this, we need to identify the locations of wavefronts. 

The geometry with a single localized shockwave was studied in detail in \cite{Shenker:2013pqa}. In that paper, the authors considered the case of a perturbation entering at a very early time so the corresponding shockwave lies on the horizon at $u = 0$. For our purpose, in order to see the initial growth of the post-collision region, we need to consider a perturbation entering at finite boundary time. 

We assume a localized perturbation propagating into the bulk from the left boundary at time $t_{wL}$ and spatial location $x_{wL}$. Where is the shockwave? The null plane given by $u = e^{t_{wL}}\neq 0$ is not the right answer as the shockwave has to stay inside its causal light cone. So away from the source the shockwave effectively comes in at a later time. A naive guess for the amount of delay would be that at spatial location $x$, the shock is delayed by the amount $\frac{|x-x_{wL}|}{v_B}$; this, however, is not correct. In fact, the delay is controlled by the speed of light: at transverse location $x$, the perturbation effectively enters at time $t_{wL}+|x-x_{wL}|$. Notice that this is the boundary of the causal light cone in the circuit picture (figure \ref{fig:localizedIntro2}). We will now justify this, first analytically, then numerically.

\begin{figure}
 \begin{center}                      
      \includegraphics[width=3.4in]{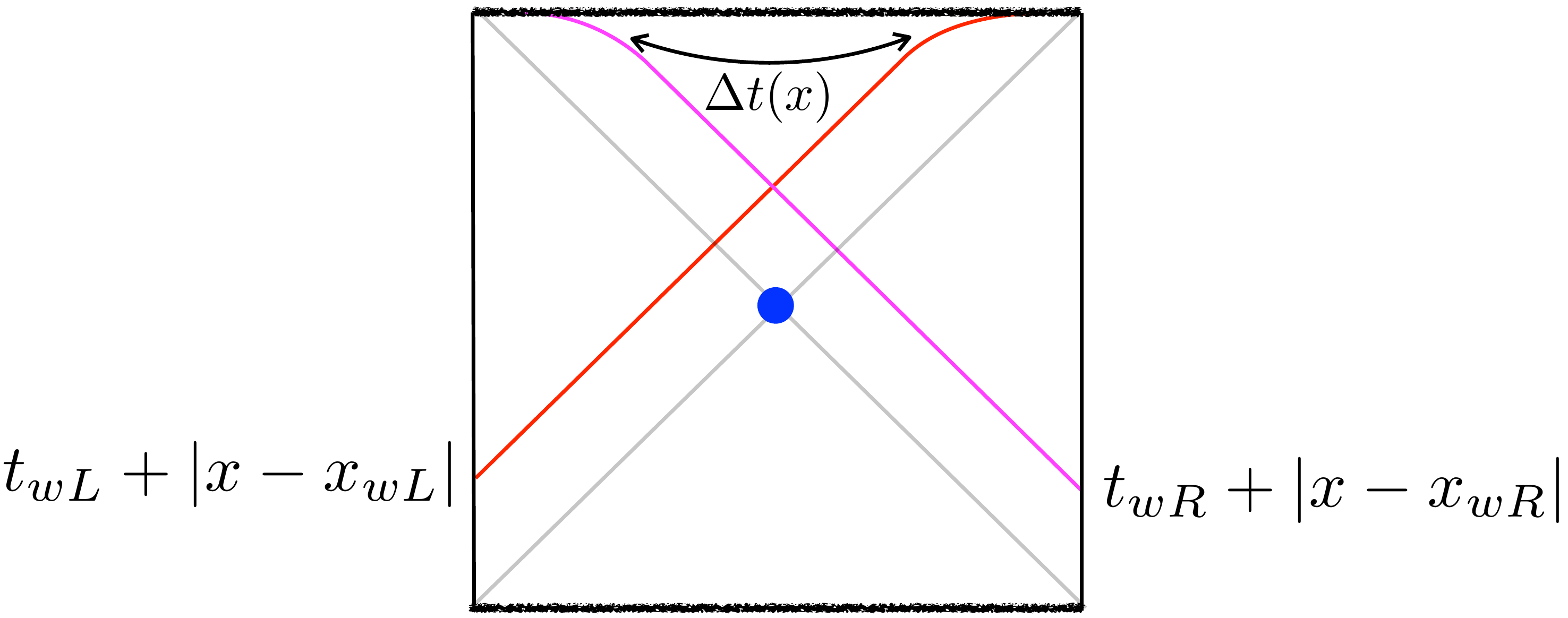}
      \caption{Localized shocks intersect the boundary at a location controlled by the causal light cone, i.e., the speed of light as opposed to the butterfly velocity. To determine the location of the shockwave, we can thus effectively take it to originate from a later time (shifted by $|x-x_{wL}|$ etc.). For $x$ away from the sources, the shocks deviate from null lines close to the singularity.}
  \label{regime_1_localized}
  \end{center}
\end{figure}

To justify the above assertion, we consider the examples of planar BTZ or Rindler-AdS in higher dimensions. In these examples we can work out the location of the shockwave hypersurface explicitly. We provide details in appendix \ref{app:location}. We consider the intersection of the shockwave with a constant $x$ slice (equation \eqref{surface_translated} where $x_{wL}$ was taken to be $0$). At the source $x = x_{wL}$ the intersection is given by a null line $u = e^{t_{wL}}$. Away from source it is no longer an exact null line, but almost, in the sense that $\log(u)-t_{wL}\approx |x-x_{wL}|$. This approximation is good until close to the singularity. In fact, from observing \eqref{surface_translated}, one can see that the intersection of the shockwave with the boundary is exactly the causal light cone given by $\log(u)-t_{wL} = |x-x_{wL}|$, while the interaction of the shockwave with the horizon $v = 0$ is given by $\log(u)-t_{wL} = \log(\cosh(|x-x_{wL})|)$. This is illustrated in figure \ref{regime_1_localized}.

Since the metric of planar BTZ restricted to a constant $x$ slice is the same as that of spherically symmetric BTZ, we can employ our previous argument regarding the spherically symmetric case. Indeed, making the approximation that the signal enters from the boundary at a time delayed by the distance from the source (figure \ref{regime_1_localized}), we can conclude that the post-collision region grows linearly in $\Delta t(x)$ at early times.
To justify this approximation, we numerically computed the spacetime volume in the post-collision region as a function of $x$ and $t$ (see appendix \ref{app:location}).

\begin{figure} 
 \begin{center}                     
      \includegraphics[width=\textwidth]{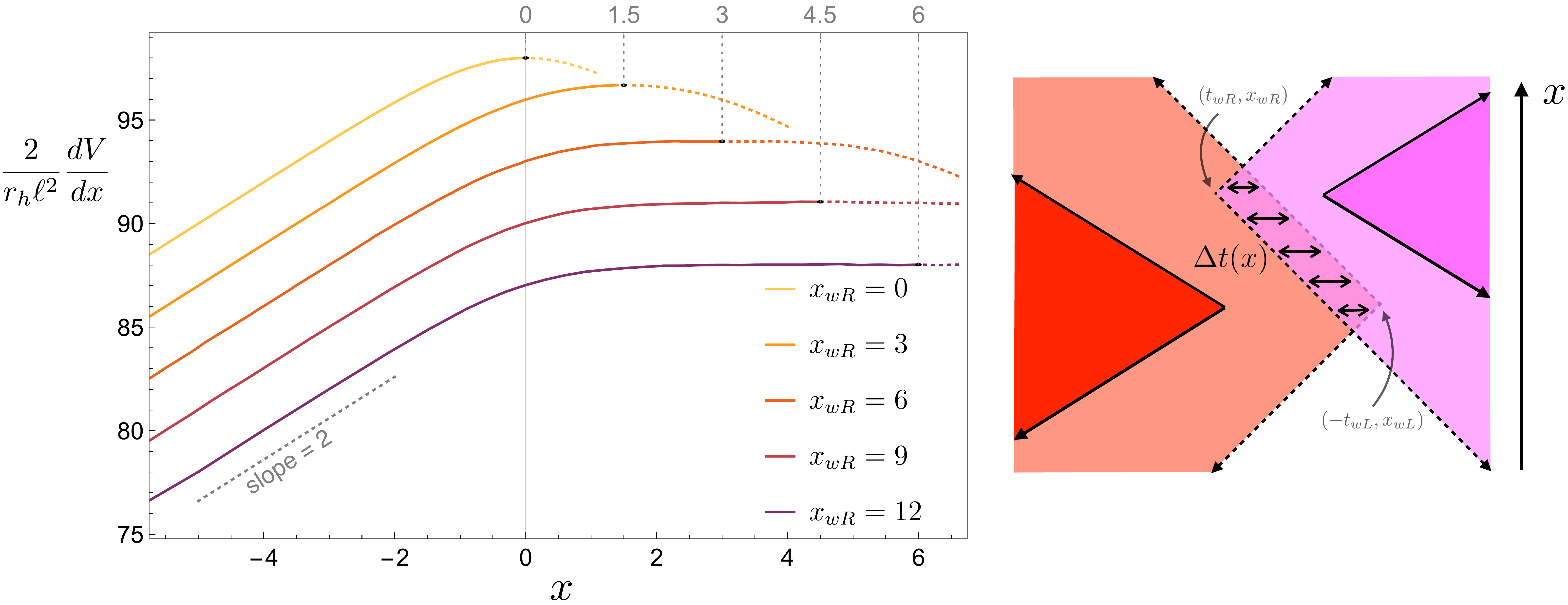}\vspace{-.3cm}
      \caption{{\it Left:} Numerical results for the area of constant $x$ slices of the post-collision region in planar BTZ geometry. The fixed parameters are $(-t_{wL},x_{wL}) = (50,0)$ and $(t_{wR},x_{wR}) = (-50,\,0\cdots 12)$. The growth with constant slope for negative $x$ is a reflection of causality. For larger $x_{wR}$ the transition to an extended constant plateau is evident. Curves are reflection symmetric around the locations indicated on the top edge of the plot. {\it Right:} Corresponding quantum circuit picture. The form of the lightcones gives $\Delta t$ an $x$-dependence that is in excellent correspondence with the rather non-trivial gravity result.}
  \label{regime_a_compare}
  \end{center}
\end{figure}

We first fix $t$ and consider the volume of the post-collision region as a function of $x$. The right panel of figure \ref{regime_a_compare} shows the circuit picture in the early time regime where $N_{\text{healthy}}(t)\propto \Delta t(x)$. We see that with a non-zero impact parameter, as we change $x$, $\Delta t(x)$ first increases linearly with $x$, and then stays constant, and eventually decreases linearly. The linear increase and decrease happen at the same rate. The left panel of figure \ref{regime_a_compare} shows corresponding numerical results in gravity:\footnote{ In order to obtain the results, we computed the area of constant $x$ slices in planar BTZ numerically. The post-collision region of interest is bounded by three curves: on the one hand, the singularity at $uv=1$, and, one the other hand, the intersection surfaces of shockwaves given by \eqref{surface_translated} and \eqref{surface_translated2}.} we plot the volume of constant $x$ slices of the post-collision region in the planar BTZ geometry. We fix $t_{wL}, x_{wL}, t_{wR}$, and draw multiple curves corresponding to different value of $x_{wR}$. We see linear growth with $x$, and with large enough impact parameter (controlled by $x_{wR}$) we see the transition to an extended constant plateau value. Curves are reflection symmetric around the indicated points. This gravity result exhibits smooth transitory behavior, but is otherwise precisely of the form predicted by the circuit picture. We note that, while these effects are obvious to see in the circuit picture, they are highly non-trivial from the bulk point of view.

\begin{figure} 
 \begin{center}                     
      \includegraphics[width=1\textwidth]{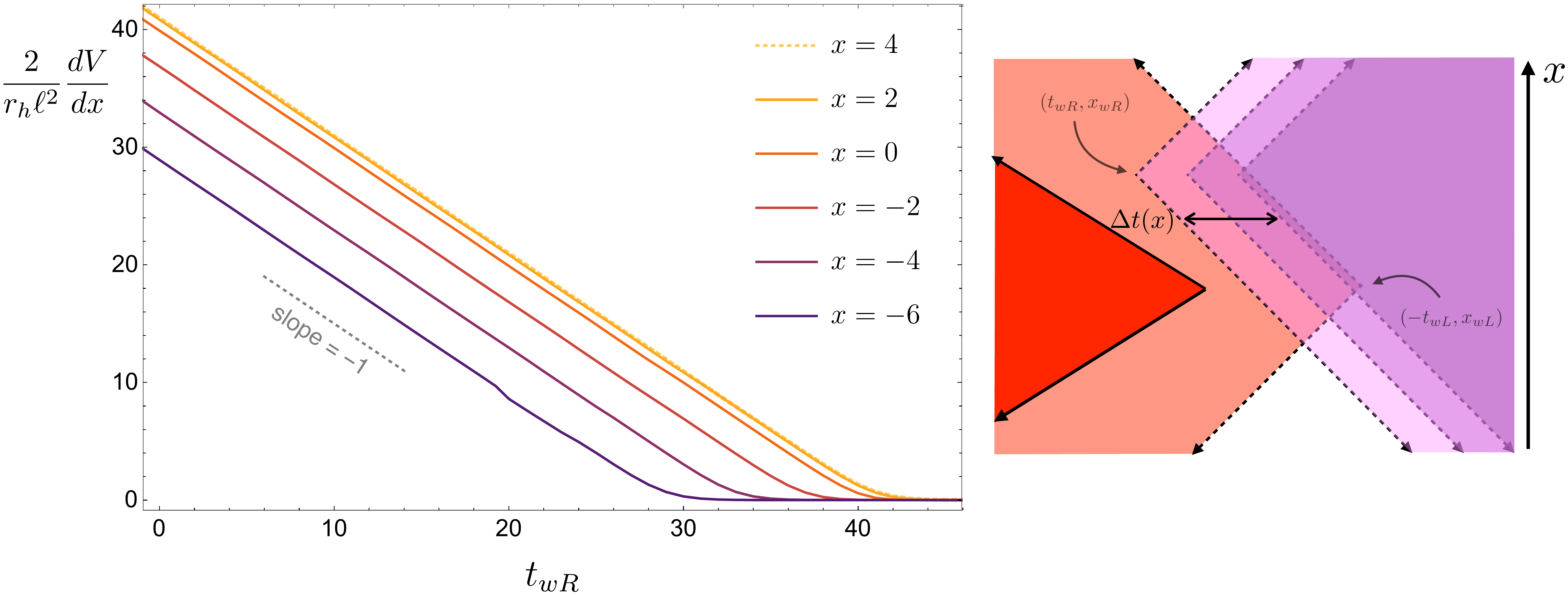}\vspace{-.3cm}
      \caption{{\it Left:} Numerical results for the area of a constant $x$ slice of the post-collision region in planar BTZ geometry. The fixed parameters are $(-t_{wL},x_{wL}) = (50,0)$, $x_{wR} = 9$, and $x=-6,\ldots,4$. The  growth with constant slope is a reflection of causality. The convergence of curves with $x_{wL}\lesssim x \lesssim x_{wR}$ to a limiting curve is due to the plateau seen in figure \ref{regime_a_compare}. {\it Right:} Corresponding quantum circuit picture.}
  \label{regime_a_compare2}
  \end{center}
\end{figure}

Next, we fix $x$ and consider the volume of the post-collision region as a function of time. In figure \ref{regime_a_compare2}, we fix $t_{wL}, x_{wL}, x_{wR}$ and vary $t_{wR}$. We draw multiple curves corresponding to different value of $x$. We clearly see the linear decrease with $t_{wR}$, which is expected from the circuit picture. The slope with respect to $t_{wR}$ here is one half of the slope with respect to $x$ in figure \ref{regime_a_compare}. This is consistent with the circuit prediction that $N_{\text{healthy}}(x)$ is proportional to $\Delta t(x)$ where $\Delta t(x) = -t_{wL}-t_{wR}-|x-x_{wL}|-|x-x_{wR}|$. This linear expansion of the post-collision region with $\Delta t(x)$ is the first regime we discussed before (e.g., figure \ref{fig:localizedSummary}(a)). Finally, note that figure \ref{regime_a_compare2} demonstrates a limiting curve as we increase $x \rightarrow x_{wR}$. This convergence of curves is equivalent to the extended plateau visible in figure \ref{regime_a_compare}.

\subsubsection{Shrinking of the post-collision region}

Let us now understand the later regimes, i.e., the shrinking of the post-collision region due to backreaction. The metric with a single shockwave exhibits a shift along the horizon, which is now $x$-dependent \cite{Dray:1985yt,Sfetsos:1994xa,Shenker:2013pqa}. The amount of shift $h(x)$ was worked out in \cite{Roberts:2014isa} for planer AdS black holes and in \cite{Ahn:2019rnq} for hyperbolic AdS black holes. The metric generally takes the following form:\footnote{ Via a coordinate transformation $v \rightarrow v - \theta(u)h(x)$, this parametrization of the metric is equivalent to one without any step functions and instead a $\delta$-function localized contribution $g_{uu}du^2 = A(uv)h(x) \delta(u) \ell^2 du^2$.}
\begin{align}
    \label{localized_BTZ_0}
    ds^2 = \ell^2\big[- A\big(u(v+\theta(u)h(\vec x))\big) \, (dudv+\theta(u)\nabla h(\vec x)\cdot dud\vec x)+ B\big(u(v+\theta(u)h(\vec{x}))\big) \, d\vec x^2\big] \,,
\end{align}
where $d\vec x^2$ is replaced by the line element for hyperbolic space, $d\mathbb{H}_{d-1}^2$, in the case of hyperbolic AdS black holes.
We first consider 2+1 dimensions, where
\begin{align}
\label{AB}
    A(x) = -\frac{4}{(1+x)^2} \, ,\qquad B(x) = \frac{r_h^2}{\ell^2}\qty(\frac{1-x}{1+x})^2 \,.
\end{align}
In order to understand the  geometry with two colliding shockwaves, let us begin with some qualitative observations. We focus on the quantity $h_1(x)h_2(x)$,\footnote{We thank Douglas Stanford for emphasizing this to us.} where $h_i(x)$ is the shift along the horizon. It is straightforward to show that this combination can be written as \cite{Roberts:2014isa,Ahn:2019rnq}
\begin{align}
	h_1(x)h_2(x) \approx e^{\Delta t(x)-t_{*1}(x)-t_{*2}(x)}\,.
\end{align}
Note that the definition of $\Delta t(x)$ in \eqref{eq:DeltatDef} does not involve the butterfly velocity, while $t_{*i}(x)$ does. 

On the circuit side, the quantity $h_1(x)h_2(x)$ characterizes the amount of overlap of the perturbations at location $x$. In gravity, it controls the strength of non-linear gravitational effects. The circuit model suggests what should occur in gravity as a function of $\Delta t(x)$: it implies that the singularity will bend down and the shared interior region becomes small at transverse locations $x$ where $h_1(x)h_2(x)$ becomes of order $1$. This is reasonable because it is exactly when the non-linear effect of general relativity becomes important. In the following we provide arguments that this expectation is indeed borne out in gravity.

As the geometry in the post-collision region is generally unknown, we will follow the same estimation method as in section \ref{sec:estimate_Raychaudhuri_1}. 
To understand the effect of a single shock, we consider again a family of null lines generated by the null vectors $K$, where
\begin{align}
    (K^u, K^v, K^x) = \big(C_1(1+u(v+h(x)\theta(u)))^2,\,0,\,0\big)\,.
\end{align}
The setup is the same as that illustrated in figure \ref{Raychaudhuri_1}, but with a $x$-dependent horizon shifts $h(x)$ and a shockwave effectively entering at a later time away from the source.
See appendix \ref{app:localized_3D} for more details. 

We apply Raychaudhuri's equation to this congruence of lines and focus on the surface of the shockwave $u = 0$. 
Recall that in section \ref{sec:estimate_Raychaudhuri_1} we observed a jump of the expansion across the shockwave due to the matter stress-energy tensor. In the present case, the stress-energy tensor is zero away from the source $x_{wL}$. However, we still observe a jump on the expansion, since now the last term in the Raychaudhuri equation \eqref{Raychaudhuri} has a delta-function contribution across the shockwave away from the source. The reason is that with $x$-dependent shift $h(x)$ the radial null lines generated by $K$ are no longer geodesics:
\begin{align}
    &\dot K^a_{\;;a} = -\frac{2C_1^2\ell^2}{r_h^2}h''(x)\delta(u)= -2C_1^2h(x)\delta(u)+8\pi G_N C_1^2T_{uu}
\end{align}
where in the second equality we used the equation of motion satisfied by $h(x)$ such that the metric \eqref{localized_BTZ_0} solves Einstein's equations \cite{Dray:1984ha,Sfetsos:1994xa,Shenker:2013pqa}. The combination appearing in Raychaudhuri's equation \eqref{Raychaudhuri} can thus be written as follows:
\begin{align}
\label{two_term}
\dot K^a_{\;;a}    -R_{ab}K^aK^b = \left[-2C_1^2h(x)\delta(u)+8\pi G_N C_1^2T_{uu} \right]-8\pi G_N T_{uu}K^uK^u = -2C_1^2h(x)\delta(u)
\end{align}
This implies that across the shockwave, the expansion jumps from $0$ to a negative value $\theta_1 = -2C_1h(x)$. Note that the jump of the expansion has exactly the same form as in the spherically symmetric case in section \ref{sec:estimate_Raychaudhuri_1}, even though the underlying physical reasons are different. This further puts an upper bound on the distance to the singularity:
\begin{align}
\label{bound_3}
    L_1(x)\leq\frac{1}{2C_1h(x)} \,.
\end{align}
With two shockwaves present as in the right panel of figure \ref{Raychaudhuri_1}, and using again the normalization $K_1\cdot K_2 = -2\ell^2$, we obtain a physical bound
\begin{align}
\label{bound_4}
    L_1(x)L_2(x)\leq\frac{1}{4h_1(x)h_2(x)}\,.
\end{align}

In $d+1=3$ spacetime dimensions, the geometry is required to be locally $AdS_3$. We thus take the following ansatz for the metric of post-collision region:
\begin{align}
\label{ansatz_3D}
	ds^2 = -\frac{4\ell^2 du dv}{(1+uv)^2}+\tilde r_h^2(x)\qty(\frac{1-uv}{1+uv})^2 dx^2 \,,
\end{align}
where $\tilde{r}_h(x)$ is a parameter of the solution that will be determined later.\footnote{It is a special feature in $2+1$ dimensions that the solution remains locally $AdS_3$ for an arbitrary function $\tilde r_h(x)$.}
One can show that the geodesic equation is the same as in the spherically symmetry case once we impose $x = \text{const}$. A computation completely analogous to subsection \ref{subsec:post_collision} gives 
\begin{align}
\label{tilde_rh_localized}
   \frac{\tilde{r}_h(x)}{r_c(x)} = \sqrt{1+\frac{1}{L_1(x)L_2(x)}}\,.
\end{align}

Combining \eqref{tilde_rh_localized}, \eqref{bound_volume_0} and \eqref{bound_4}, we get an upper bound on the spacetime volume in the post-collision region:
\begin{equation}
\label{bound_volume_2}
\begin{aligned}
\frac{2}{ r_h\ell^2}\frac{dV}{dx} =  \ &\frac{\tilde r_h(x)}{r_h}\qty[\log(\frac{\frac{\tilde r_h(x)}{r_c(x)}+1}{\frac{\tilde r_h(x)}{r_c(x)}-1})-2\frac{r_c(x)}{\tilde r_h(x)}] 
\\
    \leq \ &\tanh(\frac{\Delta t(x)}{2})\sqrt{1+4h_1(x)h_2(x)}\,\log(\frac{\sqrt{1+4h_1(x)h_2(x)}+1}{\sqrt{1+4h_1(x)h_2(x)}-1})-2\tanh(\frac{\Delta t(x)}{2})\\
    \approx \ &\begin{cases}
 	t_{*1}(x)+t_{*2}(x)-\Delta t(x) &\quad  \Delta t(x)<t_{*1}(x)+t_{*2}(x)\\
 	\frac{2}{3}e^{-(\Delta t(x)- t_{*1}(x)-t_{*2}(x))} &\quad \Delta t(x)> t_{*1}(x)+t_{*2}(x)
 \end{cases}
\end{aligned}
\end{equation}
which is derived in complete analogy with \eqref{bound_volume_1}.
This estimation matches regimes 3 and 4 of the quantum circuit picture analysis well (e.g., figure \ref{fig:localizedSummary}(c) and \ref{fig:localizedSummary}(d)). Before we move on, we should comment on the ansatz \eqref{ansatz_3D} we use. It is well known that it can be a tricky problem to solve for the post-collision geometry after localized collisions. Certain matching conditions at the collision point have to be obeyed, which were referred to as `generalized Dray-'t Hooft-Redmount (DTR) relations' \cite{Barrabes_1990}. The matching condition we used here (continuity of the product of proper affine distances across the shockwave) turns out to be equivalent to the generalized DTR relations. The bound we obtained is only approximate simply due to the somewhat unjustified assumption \eqref{ansatz_3D}. One can show, however, that with this ansatz, Einstein's equations can be solved including the gluing conditions along the shockwaves, perturbatively to linear order in $h_i(x)$. The corrections are proportional to $h_1(x)h_2(x)$, which remains small (less than order one) for the regime we are interested in.

\subsubsection{Post-collision region in all four regimes}
\label{sec:four_regime}

As mentioned earlier, we can improve the result in \eqref{bound_volume_2} by improving the bound  \eqref{bound_4}. In obtaining \eqref{bound_4} we assumed that the shockwaves are located at the horizons $u = 0$ and $v=0$, which is only true for the late regimes. Let us now drop this assumption, and instead consider a shockwave coming in at finite time $t_{wL}$ at location $x_{wL}$. Away from the source, the shockwave is approximately located at $\log u\approx \log(u_0(x))\equiv t_{wL}+ |x-x_{wL}|$ (c.f., figure \ref{regime_1_localized}). 
The expansion of radial null vectors $K_1$ as in figure \ref{Raychaudhuri_1} is given by $-2C_1v\frac{r_h}{r_c}$ before the shockwave. The expansion jumps across the shockwave by an approximate amount $-2C_1h_1(x)\frac{r_h}{r_c}$ and is afterwards given by $\theta_1 \approx -2C_1 (v+h_1(x))\frac{r_h}{r_c}$.

With two shockwaves at locations $ u \approx u_0(x)$ and $ v \approx v_0(x)$, the product of expansions after the collision takes the form
\begin{align}
    \theta_1\theta_2 \approx\ & 4C_1C_2\frac{r_h^2}{r_c^2}\qty(v_0(x)+h_1(x))(u_0(x)+h_2(x))\\
    =\ & 4\frac{1}{(1-u_0(x)v_0(x))^2}\qty(v_0(x)+h_1(x))(u_0(x)+h_2(x))\label{product_expansion_finite} \,.
\end{align}
where in obtaining \eqref{product_expansion_finite} we again used the normalization $K_1\cdot K_2 = -2\ell^2 C_1C_2(1+u_0(x)v_0(x))^2 = -2\ell^2$.

If we compare \eqref{product_expansion_finite} with earlier calculations, we note that the crucial change here is a non-zero value of $u_0(x)$ and $v_0(x)$. They also satisfy
\begin{equation}
\begin{aligned}
    &u_0(x)h_1(x) = e^{t_{wL}+|x-x_{wL}|} \times \frac{\delta S_1}{c}e^{-t_{wL}-\frac{|x-x_{wL}|}{v_B}} = e^{-t_{*1}(x)} \,,\qquad v_0(x)h_2(x) = e^{-t_{*2}(x)}.
\end{aligned}
\end{equation}

We again obtain a bound on the product of affine distances along the horizons, which is however considerably more accurate than the previous bound (e.g., \eqref{tilde_rh_localized}):
\begin{align}
\label{product_length_finite}
    L_1L_2\leq\frac{1}{\theta_1\theta_2} &\approx \frac{(1-u_0(x)v_0(x))^2}{4(v_0(x)+h_1(x))(u_0(x)+h_2(x))} \nonumber\\
    &= \frac{(1-e^{-\Delta t(x)})^2}{4\qty(e^{-\Delta t(x)}+e^{-t_{*2}(x)}+e^{-t_{*1}(x)}+e^{\Delta t(x)-t_{*1}(x)-t_{*2}(x)})}
\end{align}
Plugging this estimate into the expression for the post-collision spacetime volume \eqref{bound_volume_0}, we find (again analogous to \eqref{bound_volume_1} and \eqref{bound_volume_2}):
\begin{equation}
\label{result_3D}
\begin{aligned}
    \frac{2}{ r_h\ell^2} \frac{dV}{dx} 
\leq\ & \tanh(\frac{\Delta t(x)}{2})\sqrt{1+\frac{4\qty(e^{-\Delta t(x)}+e^{-t_{*2}(x)}+e^{-t_{*1}(x)}+e^{\Delta t(x)-t_{*1}(x)-t_{*2}(x)})}{(1-e^{-\Delta t(x)})^2}}\\
\ &\ \ \ \ \ \ \times \log(\frac{\sqrt{1+\frac{4\qty(e^{-\Delta t(x)}+e^{-t_{*2}(x)}+e^{-t_{*1}(x)}+e^{\Delta t(x)-t_{*1}(x)-t_{*2}(x)})}{(1-e^{-\Delta t(x)})^2}}+1}{\sqrt{1+\frac{4\qty(e^{-\Delta t(x)}+e^{-t_{*2}(x)}+e^{-t_{*1}(x)}+e^{\Delta t(x)-t_{*1}(x)-t_{*2}(x)})}{(1-e^{-\Delta t(x)})^2}}-1})-2\tanh(\frac{\Delta t(x)}{2})
\end{aligned}
\end{equation}
Thanks to the improved estimations, the above expression reproduces the correct behavior as in figure \ref{plot_1} for all four regimes. In fact, one can easily see the competition of four regimes analytically from the form of $\frac{1}{L_1L_2}$ written as a sum of four exponentials in \eqref{product_length_finite}.

\begin{figure}
 \begin{center}                      
      \includegraphics[width=.45\textwidth]{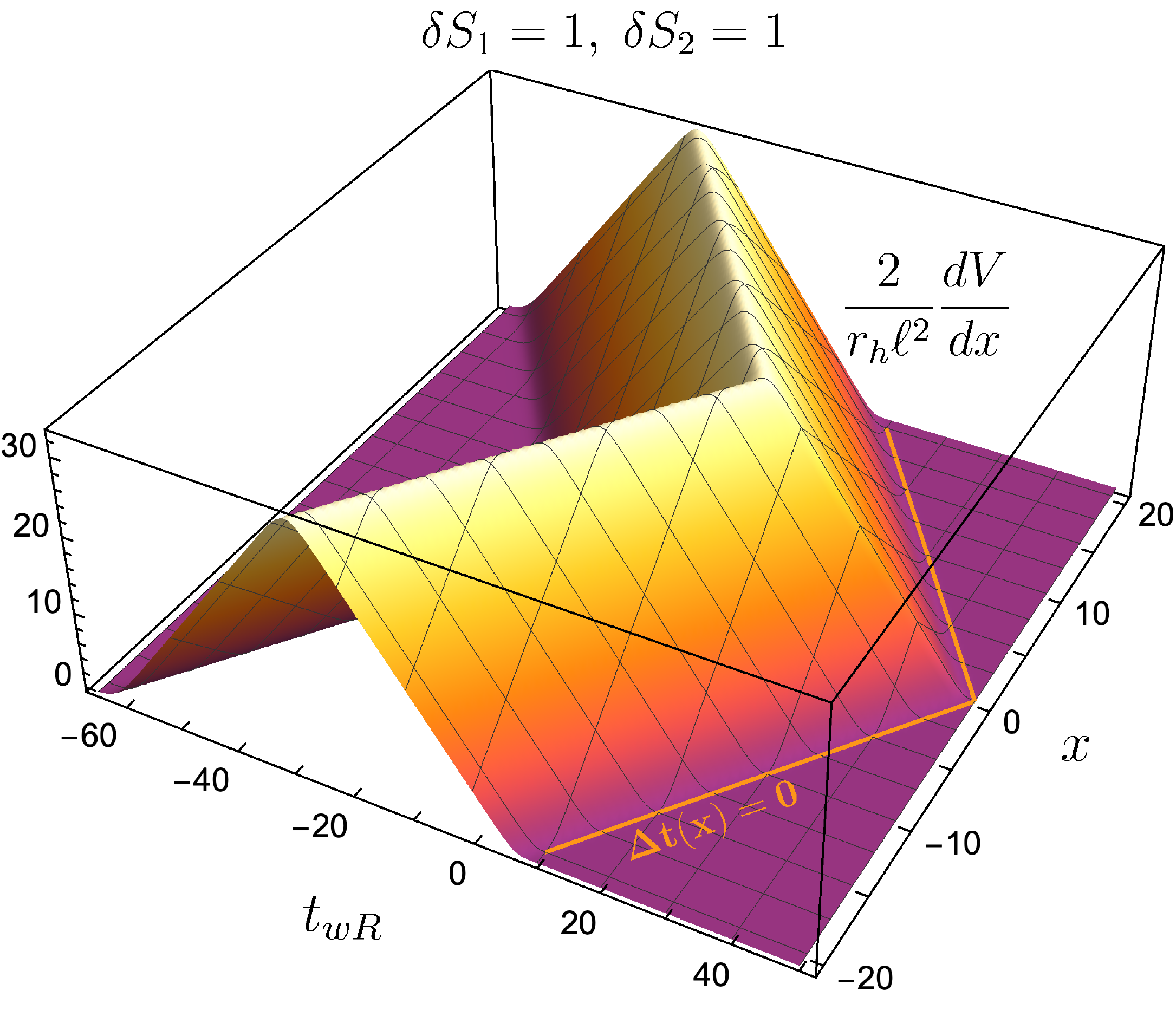}$\quad$
      \includegraphics[width=.45\textwidth]{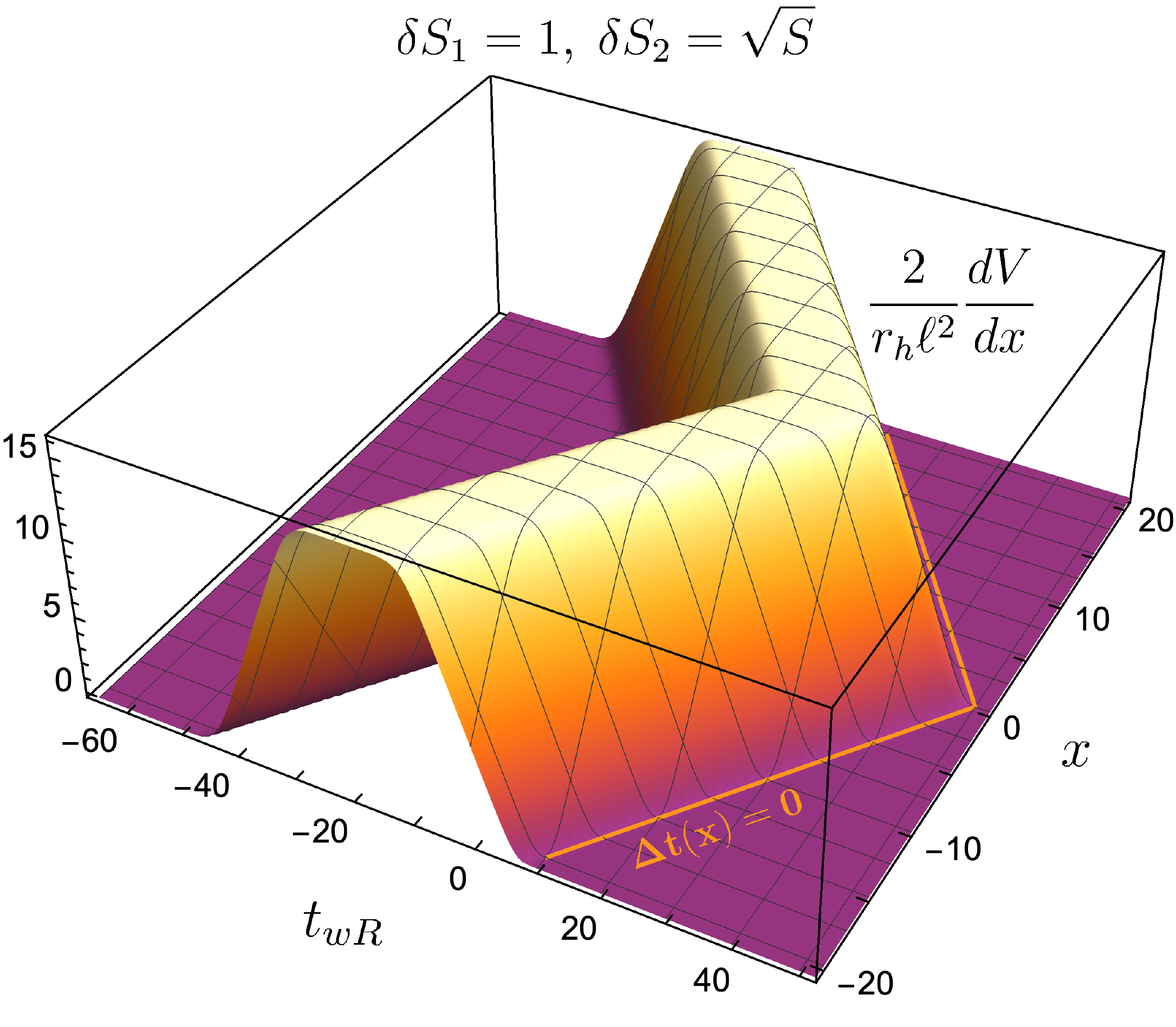}
      \vspace{-.4cm}
      \caption{Plot of the upper bound on the post-collision spacetime volume \eqref{result_3D} for head-on collisions. We set $x_{wL} = x_{wR} = 0$, $t_{wL} = -50$, $\log c = 36$ (c.f., figure \ref{plot_1}). {\it Left:} $\delta S_1 = \delta S_2 = 1$. {\it Right:} $\delta S_1 = 1$, $\delta S_2 = \sqrt S$.}
  \label{3D_12}
  \end{center}
\end{figure}

\begin{figure}
 \begin{center}                 
      \includegraphics[width=.45\textwidth]{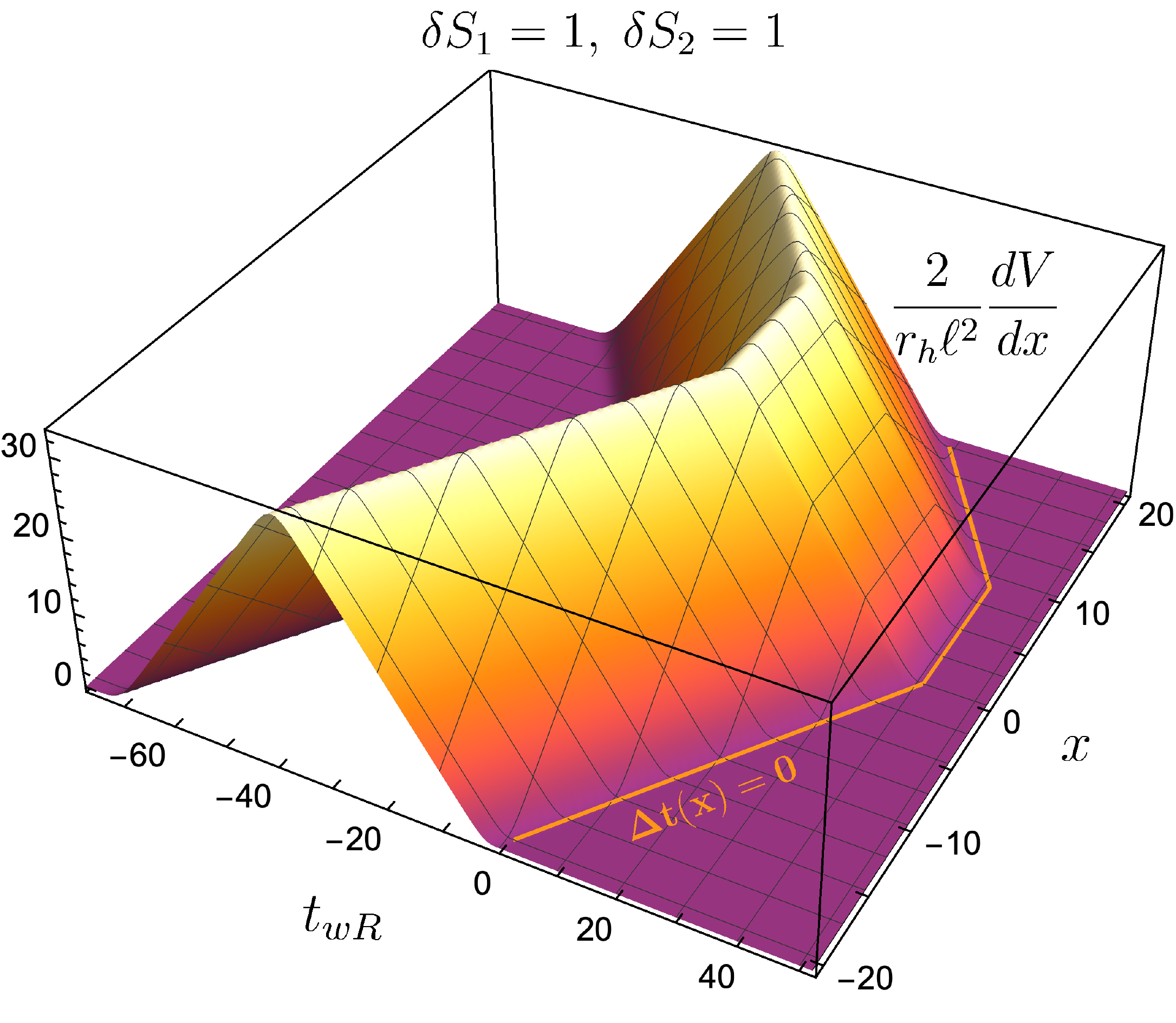}$\quad$
      \includegraphics[width=.45\textwidth]{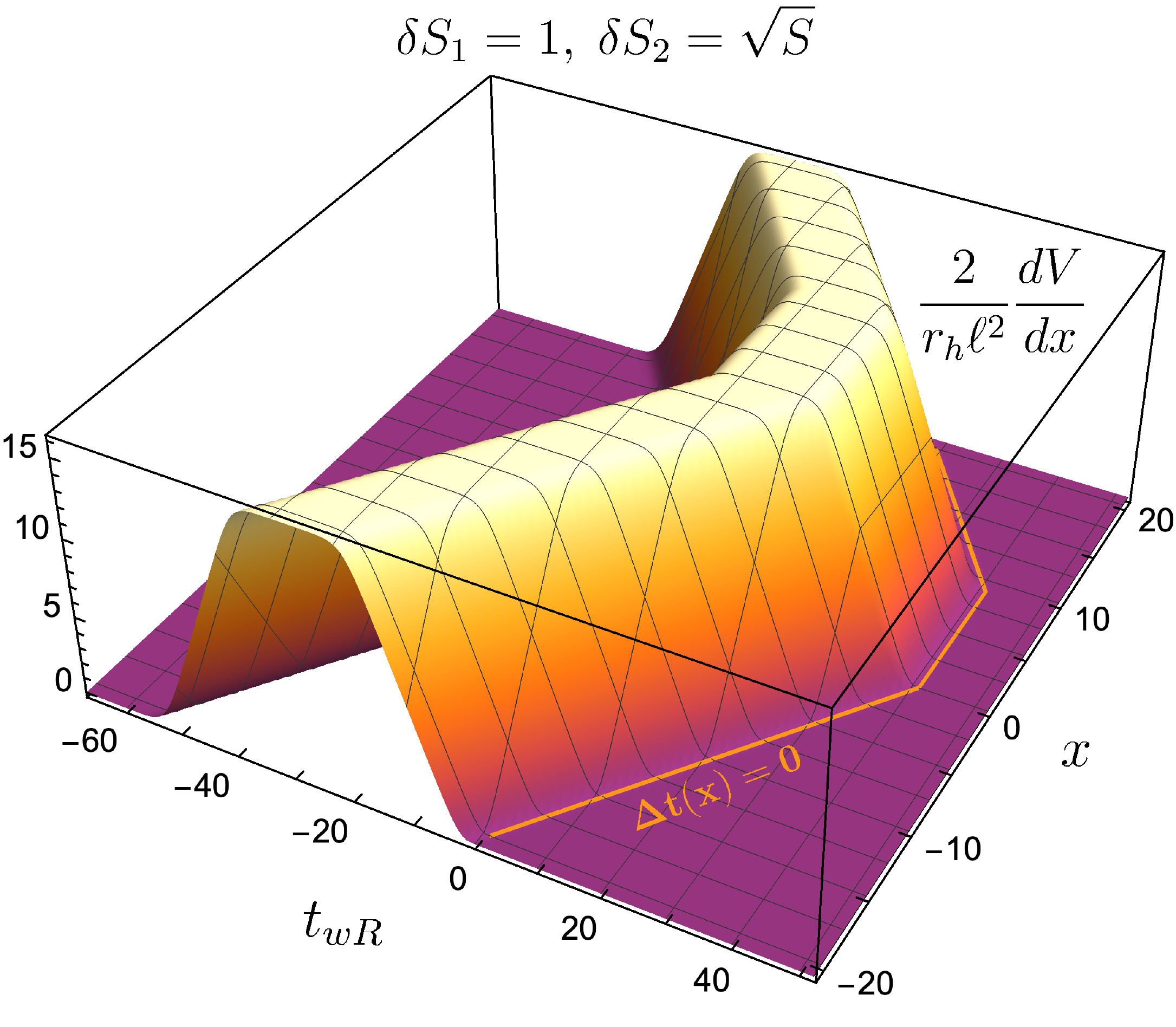}
      \vspace{-.4cm}
      \caption{Plot of the upper bound on the post-collision spacetime volume \eqref{result_3D} at finite impact parameter, where $x_{wL}= 0$, $x_{wR} = 9$, $t_{wL} = -50$, $\log c = 36$ (c.f., figures \ref{regime_a_compare} and \ref{regime_a_compare2}, which correspond to numerical results at early times for constant $t_{wR}$ and constant $x$, respectively). {\it Left:} $\delta S_1 = \delta S_2 = 1$. {\it Right:} $\delta S_1 = 1$, $\delta S_2 = \sqrt S$.}
  \label{3D_34}
  \end{center}
\end{figure}

The result \eqref{result_3D} applies when $\Delta t(x)\gg 1$. Using the definition of $\Delta t(x)$ in \eqref{eq:DeltatDef} as well as the scrambling times \eqref{eq:tstarDef}, we can produce plots of the post-collision volume as a function of both $x$ and $t$, and for any impact parameter.
Specifically, figures \ref{3D_12} and \ref{3D_34} show the dependence of the upper bound on the post-collision volume $\frac{dV}{dx}$ as a function of $x$ and $t_{wR}$, with $t_{wL}$, $x_{wL}$ and $x_{wR}$ fixed. For constant $x$ slices, we observe the same shape as for the exact results in figure \ref{plot_1} with the location of peaks shifting with $t_{wR}$. The upper bound \eqref{result_3D} therefore actually provides a very accurate description of the exact behavior.
We also see excellent qualitative agreement with the numerical results at early times for fixed $t_{wR}$ (figure \ref{regime_a_compare}) and fixed $x$ (figure \ref{regime_a_compare2}).

\subsubsection{Comments on higher dimensions}

So far we did computations in $d+1 = 3$ spacetime dimensions. In this section, we briefly consider AdS-Rindler spacetimes in higher dimensions, which are hyperbolic black holes \cite{Roberts:2014isa,Ahn:2019rnq}.\footnote{ See also \cite{Perlmutter:2016pkf,Mezei:2019dfv,Haehl:2019eae} for related recent studies.} The unperturbed metric is given by
\begin{align}
    ds^2 = -\frac{4\ell^2 dudv}{(1+uv)^2}+\ell^2\qty(\frac{1-uv}{1+uv})^2d{\mathbb H}_{d-1}^2 \,,
\end{align}
where
\begin{align}
   d{\mathbb H}_{d-1}^2 = d\chi^2+\sinh^2\chi \, d\Omega_{d-2}^2
\end{align}
is the metric of a $(d-1)$-dimensional hyperbolic plane.

In order to study the growth of the post-collision region, we need to know the location of a finite time shockwave surface. We again start from a shockwave at $u = 0$ and perform a global time translation. By the symmetry on the hyperbolic plane ${\mathbb H}_{d-1}$, we let the source be located at $\chi = 0$. By spherical symmetry in $S^{d-2}$, the location of the shockwave can only be a function of $\chi$ and therefore the coordinate transformation is exactly the same as in the 3-dimensional case. We can thus draw the same conclusion: away from the source, the shockwave effectively originates at a later time, which is delayed by an amount equal to the proper distance from the source. Ignoring backreaction, we would find for the differential spacetime volume of the post-collision region:
\begin{align}
	\frac{d}{\ell^{d+1}\text{Vol}(S^{d-2})} \frac{dV}{d\chi} =  \text{B}\qty(\frac{1-u_0(x)v_0(x)}{1+u_0(x)v_0(x)};d,0)-\text{B}\qty(\frac{1-u_0(x)v_0(x)}{2};d,1-d)\,,
\end{align}
where $u_0(x)v_0(x) = e^{-\Delta t(x)}$ and $\text{B}(z;a,b)$ is the incomplete beta-function.
If we perform an expansion for small $u_0(x)v_0(x)$, which works when $\Delta t(x)$ is bigger than a few thermal times, we find
\begin{align}
	\frac{d}{\ell^{d+1}\text{Vol}(S^{d-2})}\frac{  dV}{d\chi } =\ &
	\log(\frac{1}{u_0(x)v_0(x)}) -\tilde{C} +\mathcal{O}(u_0(x)v_0(x)) \nonumber \\
	= \ & \Delta t(x) -\tilde{C} +\mathcal{O}(e^{-\Delta t(x)})
\end{align}
where the constant $\tilde{C} = H_{d-1}+\text{B}(\frac{1}{2};d,1-d)+\log(2)$ depends on the harmonic number $H_{d-1}$. We observe that for arbitrary dimensionality $d$, the post-collision region grows linearly in $\Delta t(x)$ if backreaction can be ignored. 

Next, we incorporate backreaction effects and consider the associated shrinking of the post-collision region. With one shockwave from the left boundary, the shift of the horizon is given by\ \cite{Ahn:2019rnq,Roberts:2014isa}
\begin{align}
	h(x) \sim\frac{\delta S}{c} e^{-t_{wL}-\frac{1}{v_B}|x-x_{wL}|} \,,
\end{align}
where $|x-x_{wL}|$ is the proper distance on the hyperbolic plane ${\mathbb H}_{d-1}$.\footnote{For simplicity of notation we refrain from introducing a new symbol to denote hyperbolic geodesic distance.} We again consider the expansion of a family of radial null lines. In this case, we found that right after the shockwave at the horizon, the expansion is given by
\begin{align}
\label{high_D_exact}
	\theta_1 = -2C_1(d-1)h_1(x),\ \ \ \ L_1\leq\frac{d-1}{\theta_1} = \frac{1}{2C_1h_1(x)} \,.
\end{align}
With two shockwaves on the horizon and $K_1\cdot K_2 = -2\ell^2$, we again have $L_1L_2\leq \frac{1}{4h_1(x)h_2(x)}$. 

For shockwaves originating at a finite time, \eqref{high_D_exact} is replaced by
\begin{align}
\label{bound_5}
	\theta_1 \approx -2C_1(d-1)(v_0(x)+h_1(x))\frac{\ell}{r_c},\ \ \ \ L_1\leq\frac{d-1}{\theta_1} \approx \frac{1}{2C_1(v_0(x)+h_1(x))}\frac{r_c}{\ell} \,.
\end{align}

Now we study the same quantities in the post-collision geometry. The geometry in the post-collision region is unknown in this case. Naively one may follow the reasoning around \eqref{ansatz_3D} and assume the following ansatz:
\begin{align}
\label{metric_post_higherD}
	ds^2 = -\frac{4\ell^2 du dv}{(1+uv)^2}+\tilde r_h^2(x)\qty(\frac{1-uv}{1+uv})^2 d{\mathbb H}_{d-1}^2 \,,
\end{align}
but this does not solve Einstein's equations for $d>2$. However, if we assume $\frac{\tilde r_h(x)}{\ell} \approx \sqrt{1+\frac{1}{L_1(x)L_2(x)}}$ as in \eqref{tilde_rh_localized}, then
the ansatz solves Einstein's equations approximately when $\frac{1}{L_1(x)L_2(x)}\ll1$. This is equivalent to  $h_1(x)h_2(x)\ll 1$, or $\Delta t(x)<t_{*1}(x)+t_{*2}(x)$, i.e., the errors are proportional to $h_1(x)h_2(x)$.  
With these caveats in mind, the ansatz \eqref{metric_post_higherD} yields
\begin{align}
\label{eq:higherDimRes}
	\frac{d}{\ell^{d+1}\text{Vol}(S^{d-2})}\frac{dV}{d\chi} \approx \qty(\frac{\tilde r_h(x)}{\ell})^{d-1}  \qty[\text{B}\qty(\frac{1-\tilde{u}_0\tilde{v}_0}{1+\tilde{u}_0\tilde{v}_0};d,0)-\text{B}\qty(\frac{1-\tilde{u}_0\tilde{v}_0}{2};d,1-d)] \,,
\end{align}
where $(\tilde u_0,\tilde v_0, x)$ are the coordinates of the collision point in the post-collision metric \eqref{metric_post_higherD}, satisfying
\begin{align}
\label{matching_condition}
    \frac{1+\tilde u_0\tilde v_0}{1-\tilde u_0\tilde v_0} = \frac{\tilde r_h(x)}{r_c} = \sqrt{1+\frac{1}{L_1(x)L_2(x)}} \,.
\end{align}
Plugging \eqref{matching_condition} into \eqref{eq:higherDimRes} and as before, we get an estimation on the volume of the post-collision region for $\Delta t(x)<t_{*1}(x)+t_{*2}(x)$:
\begin{equation}
\begin{aligned}
\label{eq:higherDimRes_result}
	\frac{d}{\ell^{d+1}\text{Vol}(S^{d-2})}\frac{dV}{d\chi} \approx \ &
	\ \tanh^{d-1}\qty(\frac{\Delta t(x)}{2})\qty(1+\frac{1}{L_1(x)L_2(x)})^{\frac{d-1}{2}} \\
	&\ \ \ \ \times\qty[\text{B}\qty(\frac{1}{\sqrt{1+\frac{1}{L_1(x)L_2(x)}}};d,0)-\text{B}\qty(\frac{1}{1+\sqrt{1+\frac{1}{L_1(x)L_2(x)}}};d,1-d)]
\end{aligned}
\end{equation}
where from \eqref{bound_5} $L_1(x)L_2(x)$ has the same upper bound as in \eqref{product_length_finite}.

\begin{figure}
 \begin{center}                      
      \includegraphics[width=.99\textwidth]{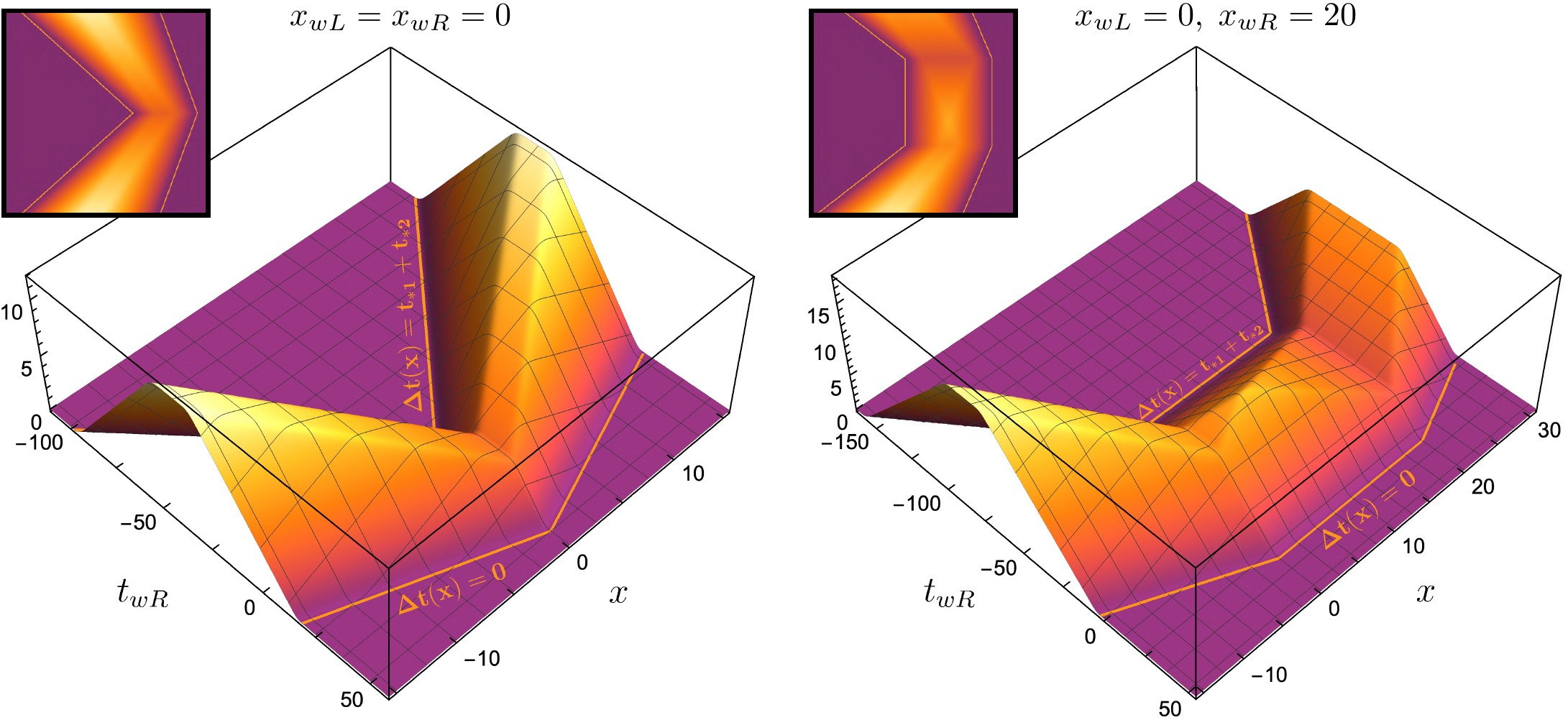}
      \vspace{-.4cm}
      \caption{Differential post-collision volume \eqref{eq:higherDimRes} in $d+1=5$ bulk dimensions as a function of $t_{wR}$ and $x$ with fixed parameters $(t_{wL},x_{wL})=(-50,0)$, $\log c = 36$. {\it Left:} Head-on collision ($x_{wR}=0$). {\it Right:} Finite impact parameter collision ($x_{wR}=20$). Insets show the view from the top; the effect of the butterfly velocity $v_B<1$ is clearly visible in both cases.}
  \label{3D_5}
  \end{center}
\end{figure}

In figure \ref{3D_5} we plot the bound from \eqref{eq:higherDimRes_result} as a function of $t_{wR}$ and $x$ in $4+1$ bulk dimensions. In the left panel of figure \ref{3D_5} (head-on collision) we labelled the two lines $\Delta t(x) = 0$ and $\Delta t(x)-t_{*1}(x)-t_{*2}(x) = 0$, which coincide approximately with the location where the volume element turns zero. As the butterfly velocity is less than the speed of light, one can clearly see that the two cones have different opening angles. In fact, the structure of the cones in left panel of figure \ref{3D_5} (top view) resembles the causal light cone and butterfly cone of the quantum circuit picture (e.g., figure 1 of \cite{Roberts:2014isa}) up to a rescaling of the $t_{wR}$-axis; but note that we obtained these plots by studying a rather different quantity, i.e., the volume of the post-collision region! The right panel of figure \ref{3D_5} shows the most general case with finite impact parameter and different initial sizes. This illustrates that both the circuit analysis and the gravity computation can become quite complicated but nevertheless they show a surprisingly good match.

\section{Discussion}
\label{sec:discussion}

We studied collisions between localized shockwaves in black hole interiors. We showed that the volume of the post-collision region matches surprisingly well with the prediction of an analysis of perturbations spreading through a quantum circuit.

It is straightforward to generalize our analysis to the case of multiple shockwaves as discussed in \cite{Shenker:2013yza,Stanford:2014jda,Roberts:2014isa}. In this context, earlier work mainly focused on the regime of large time separation ($\Delta t> 2t_*$). Our work fills this gap by giving a detailed description of the effects when $\Delta t<2t_*$. We see that the time-dependence of the post-collision region reflects various properties of the collision process. 

So far our circuit analysis relied on a simple epidemic model of the spread of perturbations with the assumption that different perturbations propagate like independent epidemics. On the boundary side a more detailed characterization of the `size' of the perturbations in terms of out-of-time-order six-point functions was proposed previously \cite{Roberts:2018mnp,Qi:2018bje,Haehl:2021prg,Haehl:2021tft}. We offer a preliminary study of such a quantity in appendix \ref{app:six_point_function}, which demonstrates qualitative agreement with circuit predictions. We will leave a more thorough analysis in two-dimensional CFT for future work. It is also worth emphasizing other proposals for the definition of operator complexity \cite{Parker:2018yvk,Magan:2018nmu,Barbon:2019wsy,Rabinovici:2020ryf,Kar:2021nbm,Caputa:2021sib} and it would be interesting to compare them in our context. One can potentially use these to diagnose more fine-grained properties of the collision in the interior.

General relativity is a set of complicated non-linear differential equations. The fact that rather non-trivial properties of its solutions can be reproduced by a simple circuit analysis is surprising. Having seen such a good match, it is tempting to attempt a derivation of gravitational interactions from the quantum circuit. However, here we encounter the question of what is the appropriate bulk quantity to consider?

We are all familiar with Newton's law characterizing gravitational forces, but this intuition does not apply in the present context as we are dealing with a high-energy collision process. We are also used to studying scattering in asymptotically flat space by means of an S-matrix, but since for our case the collision happens inside a black hole interior, asymptotic out-states do not exist and this method does not seem applicable. So far we have instead discussed the spacetime geometry itself (specifically its volume) after the collision, but this is not a conventional quantity used to characterize a scattering processes.
Can one make a connection between the post-collision geometry we studied here and more conventional characterizations of scattering processes in appropriate regimes (e.g., for relatively weak collisions)? This question is part of our motivation to study the Rinder-AdS geometry, as in this case the collision product can reach the AdS boundary.\footnote{We thank Matthew Headrick for suggesting this.} We intend to explore this in future work. As for very strong collisions, we do not know what is the correct bulk quantity to consider and maybe completely new ideas are needed. 

Some other possible future directions include the study of stringy corrections of the high energy collision process \cite{Amati:1987uf,Giddings:2007bw,Shenker:2014cwa} and the effect of electric charges \cite{Susskind:2020fiu}. For scattering in AdS, from the bulk perspective there is a transition to the formation of a black hole once the collision energy is high enough. Can this be diagnosed from the boundary?

In a broader context, there are many unanswered questions. The results of our computations mostly depend on the properties near the black hole horizon, and the black hole singularity remains a mystery. For charged black holes with inner horizons, we do not know how to characterize a collision occurring near the inner horizon when both signals are sent in late. See \cite{Lensky:2020fqf} for some recent explorations of this region. 

One may also wonder how to make the picture of the black hole interior as a unitary quantum circuit more rigorous (see \cite{Caputa:2018kdj,Erdmenger:2021wzc,Caputa:2021ori} for recent proposals in the context of CFTs), and how to connect it with other discussions of the black hole interior (such as \cite{Leutheusser:2021qhd,Leutheusser:2021frk}). Throughout this work we had a fixed Hamiltonian and the setting of AdS/CFT in mind. How to incorporate different states of baby universes if there are any \cite{Marolf:2020xie,Qi:2021oni,Almheiri:2021jwq}? We have been working to understand matter propagating in the interior in terms of a quantum circuit. Could this help us understand firewalls \cite{Almheiri:2012rt,Almheiri:2013hfa,Susskind:2015toa}?

\section*{Acknowledgements}
We thank Matthew Headrick, Gary Horowitz, Juan Maldacena, Don Marolf, Pratik Rath, Douglas Stanford, Leonard Susskind, Amirhossein Tajdini and Jieqiang Wu for helpful discussions and comments. We are especially grateful to Douglas Stanford for emphasizing the quantity $h_1(x)h_2(x)$ to us and for Matthew Headrick for suggesting us to study Rinder-AdS. 
F.H.\ acknowledges support provided by the DOE grant DE-SC0009988. 
Y.Z.\ was supported in part by the National Science Foundation under Grant No. NSF PHY-1748958 and by a grant from the Simons Foundation (815727, LB).

\appendix

\section{Location of a localized shockwave}
\label{app:location}

In this appendix we work out the location of a gravitational shockwave coming in at finite time.

We first consider planer BTZ, which is Rinder-$AdS_3$.
The location of a localized perturbation travelling into the bulk from the boundary in global $AdS_3$ is known. In order to find its form in the BTZ geometry, we transform to Rindler coordinates. Concretely, we consider the following embedding coordinates of $AdS_3$:
\begin{align}
	-T_1^2-T_2^2+X_1^2+X_2^2 = -\ell^2 \,.
\end{align}
In the following we set $\ell = 1$. 
A parametrization in terms of Kruskal coordinates is as follows:
\begin{align}
	&T_1 = \frac{v+u}{1+uv} ,\ \ \ \ T_2 = \frac{1-uv}{1+uv}\cosh x \,, \nonumber\\
	&X_1 = \frac{v-u}{1+uv},\ \ \ X_2 =  \frac{1-uv}{1+uv}\sinh x \,.
	\label{eq:TXembed}
\end{align}

\begin{figure} 
 \begin{center}                      
      \includegraphics[width=2.8in]{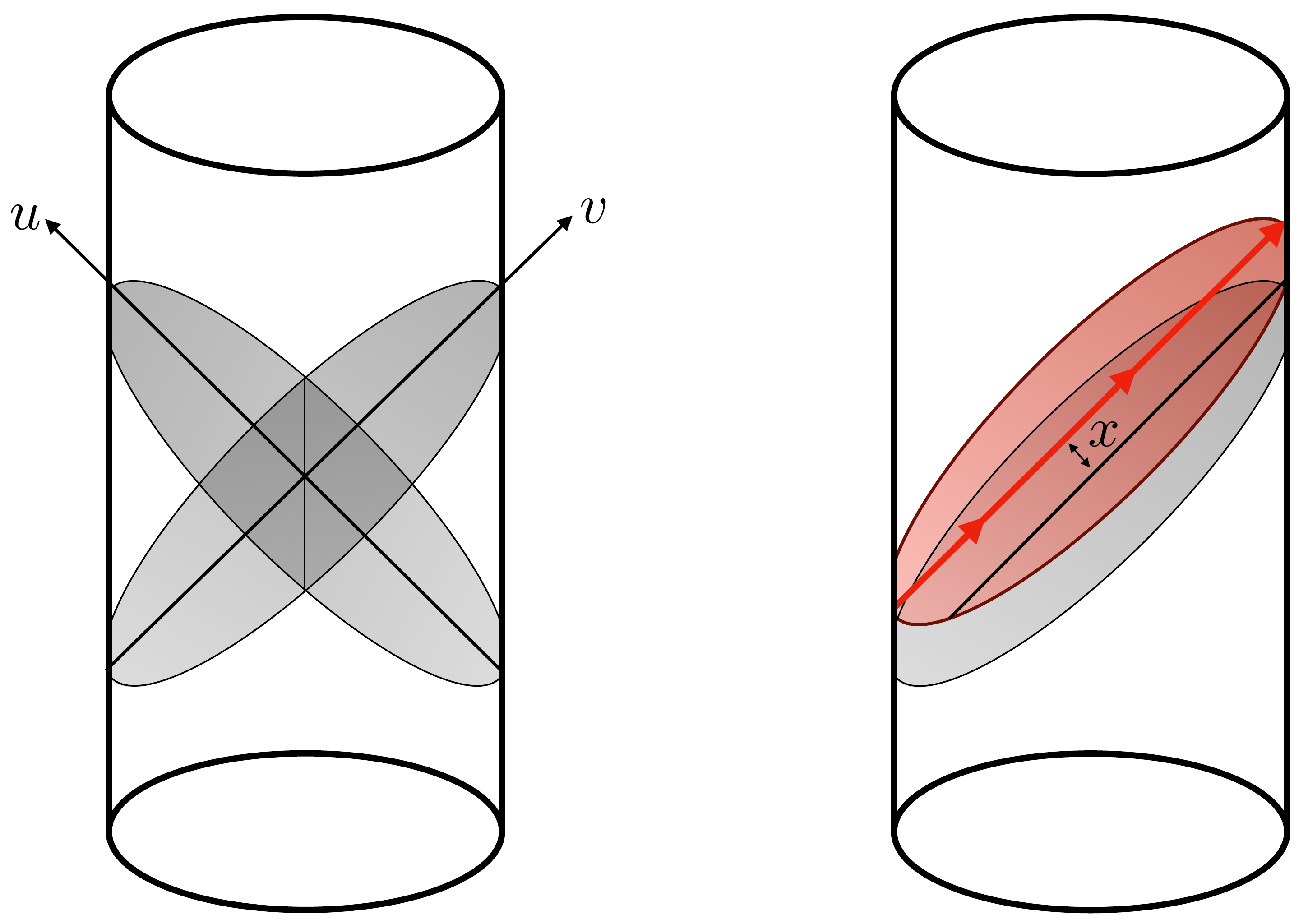}
      \caption{{\it Left:} Illustration of Kruskal-like coordinates for $AdS_3$. {\it Right:} Localized shockwave originating at finite time $t_{wL}$. The shockwave surface is related to $u = 0$ by a global time translation.}
  \label{Rindler_3D}
  \end{center}
\end{figure}

We know that $u = 0$ is one possible solution for the location of the shockwave, corresponding to a localized perturbation originating at $t_{wL} = -\infty$ and $x = 0$. The corresponding null hypersurface is parametrized by a lightcone coordinate $v$ and a transverse coordinate $x$. We wish to translate this surface by some amount of global time (figure \ref{Rindler_3D}). 
We have $\frac{T_1}{T_2} = \tan\tau = \frac{v+u}{1-uv}\frac{1}{\cosh x}$, where $\tau$ is global time. 
Under a global time translation, we observe the following transformation:
\begin{align}
\label{eq:tantauCond}
	\tau\rightarrow \tau+\epsilon\,,\quad\;\; \tan\tau\rightarrow\frac{\tan\tau+\tan\epsilon}{1-\tan\epsilon\tan\tau} \,.
\end{align}

We start from a point $(u_0,v_0,x_0)$ on the surface $u_0 = 0$, which is characterized by $X_1 = v_0$, $X_2 = \sinh x_0$, $T_1 = v_0$, $T_2 = \cosh x_0$. From \eqref{eq:TXembed} and the transformation \eqref{eq:tantauCond}, we find that the translated points $(u,v,x)$ satisfy
\begin{align}
\label{relation}
&\frac{v-u}{1+uv} = v_0\,,\quad \;\;\frac{1-uv}{1+uv}\sinh x = \sinh x_0\,,\quad  \;\; \frac{v+u}{1-uv}\frac{1}{\cosh x} = \frac{\frac{v_0}{\cosh x_0}+\tan\epsilon}{1-\tan\epsilon\frac{v_0}{\cosh x_0}} \,.
\end{align}
Combining these conditions to remove the dependence on $v_0$ and $x_0$, we see that the translated surface $(u,v,x)$ is parametrized by
\begin{align}
\label{surface_translated}
	\frac{v-u}{1+uv}\frac{1}{\sqrt{\qty(\frac{1-uv}{1+uv}\sinh x)^2+1}} = \frac{\frac{v+u}{1-uv}\frac{1}{\cosh x} -\tan\epsilon}{ \frac{v+u}{1-uv}\frac{1}{\cosh x} \tan\epsilon +1}\,.
\end{align}
At the source, $x = 0$, we have $u = \tan(\frac{\epsilon}{2}) = \text{const}$. That is, the source travels on a radial null line as expected.  We also have $u = e^{t_{wL}}$, and therefore $t_{wL} = \log(\tan(\frac{\epsilon}{2}))$. 

Similar to the derivation of \eqref{surface_translated}, one can obtain an equation describing the hypersurface of a localized shock coming from the right boundary:
\begin{align}
\label{surface_translated2}
-\frac{v-u}{1+uv}\frac{1}{\sqrt{\qty(\frac{1-uv}{1+uv}\sinh x)^2+1}} = \frac{\frac{v+u}{1-uv}\frac{1}{\cosh x} -\tan\epsilon'}{ \frac{v+u}{1-uv}\frac{1}{\cosh x} \tan\epsilon' +1}\,,
\end{align}
where $\epsilon'$ is the amount of global time translation for the second shock. Similarly, $t_{wR} = \log (\tan (\frac{\epsilon'}{2}))$.

The post-collision region is bounded by three lines described by \eqref{surface_translated}, \eqref{surface_translated2}, and the singularity at $uv=1$. It is a simple numerical task to evaluate the spacetime volume of this region as a function of the insertion points $(t_{wL},x_{wL})$, $(t_{wR},x_{wR})$. Specifically, we focus on a constant-$x$ slice and compute the post-collision volume element as a function of $x$. The results are shown in figure \ref{regime_a_compare} and figure \ref{regime_a_compare2}.

The conclusion for higher dimensional Rindler-AdS space is the same, as the coordinate transformations \eqref{eq:TXembed} are similar. We just replace $|x-x_{wL}|$ by proper distance on hyperbolic space.

\section{Geodesics on shockwave background}

\label{app:geodesic}

\subsection{Geodesics across spherically symmetric shocks}
\label{app:GeoSphSymm}

In order to derive the upper bound on affine distances from the shock to the singularity, \eqref{bound_1}, we need to understand the behavior of geodesics across shockwaves. We work in $d+1 = 3$. The metric of a single shock can be written in Kruskal form with a jump in the $v$ coordinate across $u=0$:
\begin{align}
	ds^2=\ &-\frac{4\ell^2 dudv}{[1+u(v+\theta(u)h)]^2}+r_h^2\qty(\frac{1-u(v+\theta(u)h)}{1+u(v+\theta(u)h)})^2 dx^2 \,.
\end{align}
Consider now a geodesic $(u(\tau),v(\tau),x(\tau))$ in this geometry.
The geodesic equation $\frac{D}{D\tau}K^a=0$, written in terms of its generating tangent vector $K^a(\tau)$, reads as follows: 
\begin{align}
	&\frac{d}{d\tau}K^u = \frac{2(v+h\theta(u))(K^u)^2+r_h^2 u\qty[1-u(v+h\theta(u))](K^x)^2}{1+u\qty(v+h\theta(u))}\\
	&\frac{d}{d\tau}K^v = \frac{2u(K^v)^2+r_h^2 (v+h\theta(u))\qty[1-u(v+h\theta(u))](K^x)^2}{1+u\qty(v+h\theta(u))}\\
	&\frac{d}{d\tau}K^x = \frac{4\qty[(v+h\theta(u))K^u+u K^v]K^x}{\qty[1-u\qty(v+h\theta(u))]\qty[1+u\qty(v+h\theta(u))]}
\end{align}

We consider specifically $K_1^a$ parametrizing a radial geodesic, such that $K_1^x = 0$, and the remaining equations are
\begin{align}
    &\frac{d}{d\tau}K_1^u = \frac{2(v+h\theta(u)) (K_1^u)^2}{1+u(v+h\theta(u))}\,, \qquad \frac{d}{d\tau}K_1^v = \frac{2u (K_1^v)^2}{1+u(v+h\theta(u))} \,.
\end{align}
Consider the geodesic along the null line $v = v_0$. It has $K_1^v = 0$, so the second equation is automatically satisfied. Away from $u=0$, the first equation can in turn be integrated trivially by identifying $K_1^u \equiv \frac{du}{d\tau}$:
\begin{align}
    \frac{d}{d\tau}\log(\frac{du}{d\tau}) &=  \frac{2(v_0+h\theta(u)) \frac{du}{d\tau}}{1+u(v_0+h\theta(u))} = \frac{d}{d\tau}\log(1+(v_0+h\theta(u)) u)^2\\
      \Rightarrow\qquad K_1^u &\equiv \frac{du}{d\tau} = C_1(1+u(v_0+h\theta(u)))^2
\end{align}
Note the discontinuity in derivative of $K_1^u$. We also observe that $K_{1v}$ is constant: $K_{1v} = -2\ell^2 C_1$. 

Similarly, we can determine the geodesics along $u = u_0$ across the right shockwave. We call the corresponding null vector $K_2$. Along null geodesics with $u = u_0$, we have $K_2^u = 0$ and a similar calculation as above reveals:
\begin{align}
  & K_2^v \equiv \frac{dv}{d\tau} = C_2(1+v(u_0+h_2\theta(v)))^2 \,.
 \end{align}
We compute $K_1\cdot K_2$ before the two shockwaves collide:
\begin{align}
	K_1\cdot K_2 = -2\ell^2 C_1C_2(1+u_0v_0)^2 \,.
\end{align}
In this $(2+1)$-dimensional metric, $L_1L_2$ can also be directly computed:  
\begin{align}
 	(K_1\cdot K_2)(L_1L_2) = -\frac{\ell^2}{2}\frac{(1+u_0v_0)^2}{(u_0+h_1)(v_0+h_2)}\qty(\frac{1-u_0(v_0+h_1)}{1+u_0(v_0+h_1)})\qty(\frac{1-v_0(u_0+h_2)}{1+u_0(v_0+h_2)})\,.
\end{align}
By choosing the normalizations such that $K_1\cdot K_2 = -2\ell^2$, and considering $u_0 = v_0 = 0$, we find $L_1L_2 =\frac{1}{4h_1h_2}$ which actually saturates the bound in \eqref{bound_1}.

\subsection{Geodesics across localized shocks}
\label{app:localized_3D}

We now perform a similar calculation as in the previous subsection for localized shockwaves. Let us again consider $d+1 = 3$ for simplicity. The metric with a single shock is now given by
\begin{align}
\label{metric_shock_early}
	ds^2=\ &-\frac{4\ell^2 \qty[dudv+h'(x)\theta(u)dudx]}{[1+u(v+\theta(u)h(x))]^2}+\ell^2\qty(\frac{1-u(v+\theta(u)h(x))}{1+u(v+\theta(u)h(x))})^2 dx^2\,.
\end{align}
Consider again radial null lines $v = v_0,\ x = x_0$ with generators $K^u = C_1(1+u(v_0+h(x_0)\theta(u)))^2$. These are not geodesics in the above spacetime, but the derivative of the vector $K$ vanishes once we restrict to the hypersurface $x = x_0$, i.e, $\frac{D}{D\tau}K^u$ is orthogonal to the surface $x = x_0$. In fact, a straightforward computation shows that
\begin{align}
    \qty(\frac{D}{D\tau}K_u, \frac{D}{D\tau}K_v, \frac{D}{D\tau}K_x) = \qty(0,0,-2 C_1^2\ell^2\delta(u)h'(x)) \,.
\end{align}
This also shows that this family of null lines are geodesics away from the shockwave surface. 

For shockwaves originating at finite time $t_{wL}$, we can obtain the metric by applying a global time translation to the metric describing the case of a shockwave along $u = 0$, \eqref{metric_shock_early}. The coordinate transformation will be complicated but we can work perturbatively in $e^{t_{wL}}$. This is how we obtained the approximate results for the expansion in subsection \ref{sec:four_regime}.

\section{Operator size and six-point functions}
\label{app:six_point_function}

\begin{figure} 
 \begin{center}                     
      \includegraphics[width=\textwidth]{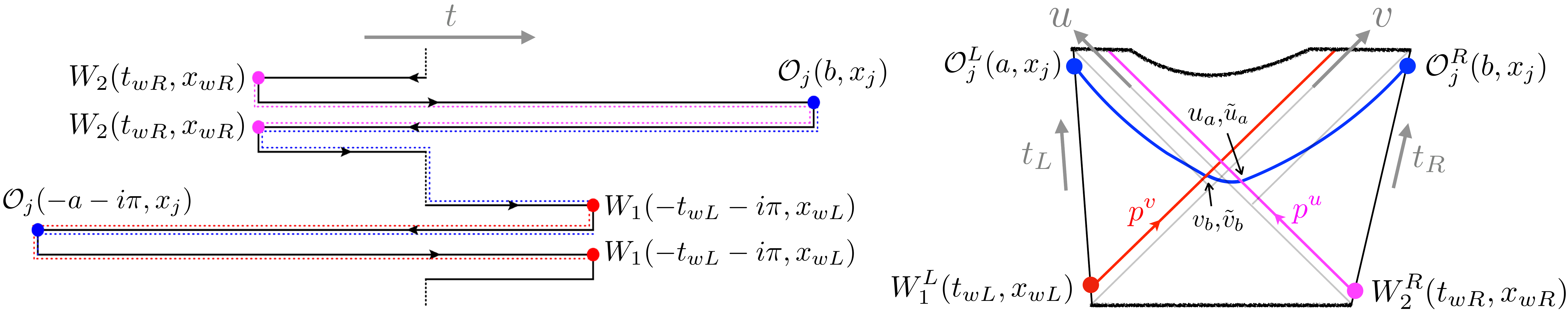}\vspace{-.3cm}
      \caption{The arrangement of operator insertions for the six-point function \eqref{eq:F6def2} along the complex time contour (left) and on the two sides of the thermofield double geometry (right). The approximate left-right geodesic is shown in blue; its interior anchor points $(u_a,v_b)$ differ from the exterior coordinates $(\tilde{u}_a,\tilde{v}_b)$ by a finite amount \eqref{eq:shift}.}
  \label{fig:6ptContour}
  \end{center}
\end{figure}

In \cite{Haehl:2021prg,Haehl:2021tft} a particular six-point function was introduced as a measure of operator size in the thermofield double state perturbed by one operator on the left and one on the right. In this section we reconsider this approach for localized perturbations in a higher-dimensional theory. The (normalized) six-point function of interest is:
\begin{equation}
\label{eq:F6def2}
  {\cal F}_6 = \frac{\big{\langle}W_1(\hat{t}_1,x_{wL}) {\cal O}_j(\hat{t}_3,x_j) W_1 (\hat{t}_2,x_{wL})  W_2(\hat{t}_5,x_{wR}){\cal O}_j(\hat{t}_4,x_j)  W_2(\hat{t}_6,x_{wR}) \big{\rangle}_\beta}{ \langle W_1(\hat{t}_1,x_{wL})W_1(\hat{t}_2,x_{wL})\rangle_\beta\, \langle {\cal O}_j(\hat{t}_3,x_j){\cal O}_j(\hat{t}_4,x_j)\rangle_\beta \,\langle W_2(\hat{t}_5,x_{wR}) W_2(\hat{t}_6,x_{wR})\rangle_\beta}
\end{equation}
where we insert pairs of identical operators at the same spatial locations $x_{wL},x_{wR},x_j$, and we defined the regularized (complex) insertion times
\begin{equation}
\label{eq:ConfigDef2}
 \hat{t}_1 = -t_{wL}- i(\pi+2\delta_1) \,,\;\; \hat{t}_2 = -t_{wL} -i\pi   \,,\;\; \hat{t}_3= -a-i\pi  \,,\;\; \hat{t}_4 = b\,,\;\; \hat{t}_5 = t_{wR}-2i\delta_2 \,,\;\; \hat{t}_6 = t_{wR}\,.
\end{equation}
The setup is illustrated in figure \ref{fig:6ptContour}. It is useful to think of ${\cal F}_6$ as a two-point function of operators ${\cal O}_j$, which probe the state $W_2^R(t_{wR},x_{wR}) W_1^L(t_{wL},x_{wL}) | \mathit{TFD}\rangle$, obtained by acting with two localized perturbations on the thermofield double (a.k.a., the two-sided purification of the thermal state).

We will assume that the operators are all scalars.
We will compute the six-point function using the eikonal formalism for gravitational shockwave scattering that was developed in \cite{HOOFT198761,Verlinde:1991iu,Kabat:1992tb,Kiem:1995iy,Cornalba:2007zb,Brower:2007qh}, introduced to the present context in \cite{Shenker:2014cwa,Maldacena:2017axo}, and adapted to six-point functions in \cite{Haehl:2021tft}. We refer especially to \cite{Shenker:2014cwa,Haehl:2021tft} for detailed explanations relevant to the calculations below.
Using the techniques developed therein, we obtain the eikonal approximation to the correlator in momentum space:
\begin{equation}
\label{eq:F6first}
\begin{split}
 &{\cal F}_6 \approx {\cal N} \int dp^udp^v dq^udq^v \int dx dx' dy dy' \, \langle W_2(x_{wR}) | p^u,y \rangle \langle p^u,y | W_2(x_{wR}) \rangle\\
 &\;\;\;\times\langle W_1(x_{wL}) | p^v,x \rangle   \langle p^v,x | W_1(x_{wL}) \rangle\;\langle   {\cal O}_j^L(x_j) {\cal O}_j^R(x_j) | q^u,x';q^v,y' \rangle\,e^{iG \left(q^up^v e^{-|x-x'|}+q^v p^u e^{-|y-y'|}\right)}
 \end{split}
\end{equation}
where $p^{u,v}$ and $q^{u,v}$ are null momenta along the horizons. 
The wave functions give a decomposition of the shockwaves in states of definite momentum. For instance, the following combinations are products of bulk-boundary propagators (i.e., Fourier transforms of boundary two-point functions): 
\begin{equation}
\label{eq:waveFc1}
\begin{split}
\langle   W_1(x_{wL}) | p^v,x  \rangle\langle p^v,x | W_1(x_{wL})\rangle
&= c_1 \, \Theta(p^v) \frac{(2p^v \, e^{t_{wL}})^{2\Delta_1}}{p^v} \, e^{-4p^v \,e^{t_{wL}} \sin \delta_1 \, \cosh (x-x_{wL})} \\
\langle   W_2(x_{wR}) | p^u,y  \rangle\langle p^u,y | W_2(x_{wR})\rangle
&= c_2 \, \Theta(p^u) \frac{(2p^u \, e^{t_{wR}})^{2\Delta_2}}{p^u} \, e^{- 4 p^u \,e^{t_{wR}} \sin \delta_2 \, \cosh (y-x_{wR})} 
\end{split}
\end{equation}
where $c_{1,2}$ are appropriate normalization constants.
Finally, we wish to express the left-right two-point function as the Fourier transform of an appropriate bulk-to-bulk propagator from a point $(u_a,v_a=0,x_a)$ on one horizon to a point  $(u_b=0,v_b,x_b)$ on the opposite horizon. The geodesic distance between two generic points is
\begin{equation}
\label{eq:geodDist}
\begin{split}
    d(u_a,v_a,x_a;u_b,v_b,x_b) &= \cosh^{-1} \left[ \frac{ (1-u_av_a)(1-u_bv_b) \cosh x_{ab}  + 2(u_av_b+v_au_b)}{(1+u_av_a)(1+u_bv_b)} \right] \,.
\end{split}
\end{equation}

We will make the following crude approximation: we write the left-right correlator as an exponential of the sum of three geodesic distances. The geodesics connect the left and right insertion points with points along the $u=0$ and $v=0$ horizons, respectively:
\begin{equation}
\begin{split}
    &\langle   {\cal O}_j^L(x_j) {\cal O}_j^R(x_j) | q^u,x';q^v,y' \rangle \\
    &\quad \approx \int du_a dv_b d\tilde{u}_a d\tilde{u}_b \, e^{i(\tilde{u}_a-u_a) q^v + i (v_b-\tilde{v}_b) q^u}\, e^{ - \Delta_j \left[ d(a,x_j ; 0,\tilde{v}_b,x') + d(0,v_b,x';u_a,0,y') + d(\tilde{u}_a,0,y'; b,x_j) \right]}
\end{split}
\end{equation}
where we explicitly wrote the Fourier transform with respect to the horizon coordinates. Geodesic distances $d$ with only five arguments mean that one of the points has been taken to the boundary (and that point is given in $(t,x)$ boundary coordinates instead of $(u,v,x)$ bulk coordinates). Specifically:
\begin{equation}
    \begin{split}
    d(a,x_j ; 0,\tilde{v}_b,x') &= \log \left[2r_c \left( \cosh(x'-x_j) +e^{a} \tilde{v}_b\right) \right]  \\ d(0,v_b,x';u_a,0,y') &= \cosh^{-1} \left[ \cosh(x'-y') +2u_a v_b \right]\\
    d(\tilde{u}_a,0,y'; b,x_j) &=\log \left[2r_c \left( \cosh(y'-x_j) +e^{b} \tilde{u}_a \right)\right]
    \end{split}
\end{equation}
where $r_c$ is the cutoff value regulating divergences in bulk-boundary propagators.

Plugging the approximate left-right correlator into the integral \eqref{eq:F6first} allows us to consider a simple saddle point approximation for $\Delta_j \gg 1$. This amounts to extremizing the total geodesic distance, which in turn leads to saddle point values for $u_a$, $v_b$, $\tilde{u}_a$, $\tilde{v}_b$, as well as $x$, $y$, $x'$, $y'$. First, we find the following shifts along the horizons:
\begin{equation}
\label{eq:shift}
    \begin{split}
    \tilde{u}_a = u_a + Gp^u e^{-|y-y'|} \,, \qquad \tilde{v}_b = v_b + Gp^v e^{-|x-x'|} \,.
    \end{split}
\end{equation}
Using these expressions, we then have saddle point values $x=x_{wL}$, $y=x_{wR}$, $x'=y'=x_j$. Finally, the saddle point values for the anchor points of the geodesics along the horizons are:
\begin{equation}
\begin{split}
    u_a^* &=  \frac{1}{2} \left[ \frac{1}{e^{-a} + Gp^ve^{-|x_{wL}-x_j|}}-e^{-b}- Gp^ue^{-|x_{wR}-x_j|} \right]  \,,\\
    v_b^* &=  \frac{1}{2} \left[ \frac{1}{e^{-b}+ Gp^ue^{-|x_{wR}-x_j|}}-e^{-a}- Gp^ve^{-|x_{wL}-x_j|}\right] \,.
    \end{split}
\end{equation}
We are left with only two integrals over the original null momenta:
\begin{equation}
\label{eq:F6first3}
\begin{split}
 {\cal F}_6 &\approx {\cal N} \int dp^udp^v \, \langle W_2(x_{wR}) | p^u,x_{wR} \rangle \langle p^u,x_{wR} | W_2(x_{wR}) \rangle\,\langle W_1(x_{wL}) | p^v,x_{wL} \rangle   \langle p^v,x_{wL} | W_1(x_{wL}) \rangle \\
 &\qquad\qquad\qquad \times e^{ - \Delta_j \left[ d(a,x_j ; 0,v_b^* + Gp^v e^{-|x_{wL}-x_j|},x') + d(0,v_b^*,x_j;u_a^*,0,x_j) + d(u^*_a+ Gp^ue^{-|x_{wR}-x_j|},0,x_j; b,x_j) \right]} \\
 &=\frac{ {\cal N}c_1c_2 }{ \left[2 \cosh \frac{a+b}{2} \right]^{2\Delta_j}} \int_0^{\infty} \frac{dp^u}{p^u} \frac{dp^v}{p^v} \, \left( 2p^v e^{t_{wL}} \right)^{2\Delta_1} \left( 2p^u e^{t_{wR}} \right)^{2\Delta_2} \, e^{-4p^v e^{t_{wL}} \sin \delta_1 - 4p^u e^{t_{wR}} \sin \delta_2 } \\
 &\quad\;\;\; \times  \, \left[1+ \frac{ Gp^v e^{\frac{a-b}{2}-|x_{wL}-x_j|}+ Gp^u e^{-\frac{a-b}{2}-|x_{wR}-x_j|} + G^2 p^v p^u e^{\frac{a+b}{2}-|x_{wL}-x_j|-|x_{wR}-x_j|} }{2\cosh \frac{a+b}{2}}\right]^{-2\Delta_j}
 \end{split}
\end{equation}
This integral is of the same form as (A.9) in \cite{Haehl:2021tft}, except for additional dependence on the transverse coordinates.
This integral can be straightforwardly evaluated by saddle point for $\Delta_{1,2} \gg \Delta_j$. We then simply find:
\begin{equation}
\label{eq:F6calc2appendix}
\begin{split}
{\cal F}_6 
   \approx
\left[ \frac{1}{1+G\Delta_1 \, z_1} \right]^{2\Delta_j} \left[ \frac{1}{1+ G\Delta_2 \, z_2} \right]^{2\Delta_j} 
\left[ \frac{1}{1+ \frac{ G^2\Delta_1\Delta_2 \, z_1 \, z_2 \, e^{a+b}}{ \left( 1+ G\,\Delta_1 \, z_1 \right) \left( 1+ G\,\Delta_2 \, z_2 \right)} }
\right]^{2\Delta_j} 
\qquad (\Delta_{1,2} \gg \Delta_j)
\end{split}
\end{equation}
where we abbreviate
\begin{equation}
\label{eq:z12pdef}
  z_1 = \frac{e^{\frac{a-b}{2}-t_{wL}-|x_{wL}-x_j|} }{8\, \sin \delta_1 \cosh \frac{a+b}{2}} \,,\qquad
  z_2 = \frac{e^{-\frac{a-b}{2}-t_{wR}-|x_{wR}-x_j|} }{8\, \sin \delta_2 \cosh \frac{a+b}{2}}  \,.
\end{equation}
For $a,b\gg -t_{wR,wL}$ it simplifies further, as we can ignore the first two factors in \eqref{eq:F6calc2appendix}:
\begin{equation}
\label{eq:F6calc2appendix2}
\begin{split}
{\cal F}_6 
   \approx
\left[ \frac{1}{1+ \frac{G^2\Delta_1\Delta_2}{16\sin\delta_1 \sin\delta_2} \, e^{-t_{wL}-t_{wR}-|x_{wL}-x_j|-|x_{wR}-x_j|}}
\right]^{2\Delta_j} 
\qquad (\Delta_{1,2} \gg \Delta_j \;,\,\; a,b \gg -t_{wR,wL})
\end{split}
\end{equation}
The physics of these expressions was discussed in \cite{Haehl:2021prg} for lower dimensions. The new feature here is the form of the $x$-dependence, but the detailed analysis in that reference otherwise applies.

%%%%%%%%%%%%%%%%%%%%%%%%%%%%%%%%%%%%%%%%%%%%%%
%%%%%%%%%%%%%%%%%%%%%%%%%%%%%%%%%%%%%%%%%%%%%%
%~~~~~~~~~~~~~~~~~~~~~~~~~~~~~~~~~~~~~~~~~~~~~~
\bibliographystyle{utphys}
\bibliography{reference}
%~~~~~~~~~~~~~~~~~~~~~~~~~~~~~~~~~~~~~~~~~~~~~~

\end{document}